\documentclass[12pt,rmp,twocolumn,natbib,showkeys]{revtex4}

\usepackage[dvips]{graphicx}

\begin{document}

\title[]
{L\'{e}vy Flight Superdiffusion: An Introduction}

\author{\firstname{A. A.} \surname{Dubkov$^{\sharp}$}}
\author{\firstname{B.} \surname{Spagnolo$^{\star}$}}
\author{\firstname{V. V.} \surname{Uchaikin$^{\flat}$}}
\affiliation{$^{\sharp}$ Radiophysics Faculty, Nizhniy Novgorod
State University \ \\23 Gagarin Ave., 603950 Nizhniy Novgorod,
Russia\footnote{e-mail: dubkov@rf.unn.ru}} \affiliation{$^{\star}$
Dipartimento di Fisica e Tecnologie Relative and CNISM-INFM,
\\ Group of Interdisciplinary
Physics\footnote{http://gip.dft.unipa.it}, Universit$\grave{a}$ di
Palermo, \\ Viale delle Scienze, I-90128, Palermo,
Italy\footnote{e-mail: spagnolo@unipa.it}} \affiliation{$^{\flat}$
Department of Theoretical Physics, Ulyanovsk State University \\
42 L.~Tolstoy str., 432970 Ulyanovsk, Russia\footnote{e-mail:
uchaikin@sv.uven.ru}}

\begin{abstract}
After a short excursion from discovery of Brownian motion to the
Richardson "law of four thirds" in turbulent diffusion, the article
introduces the L\'{e}vy flight superdiffusion as a self-similar
L\'{e}vy process. The condition of self-similarity converts the
infinitely divisible characteristic function of the L\'{e}vy process
into a stable characteristic function of the L\'{e}vy motion. The
L\'{e}vy motion generalizes the Brownian motion on the base of the
$\alpha$-stable distributions theory and fractional order
derivatives. The further development of the idea lies on the
generalization of the Langevin equation with a non-Gaussian white
noise source and the use of functional approach. This leads to the
Kolmogorov's equation for arbitrary Markovian processes. As
particular case we obtain the fractional Fokker-Planck equation for
L\'{e}vy flights. Some results concerning stationary probability
distributions of L\'{e}vy motion in symmetric smooth monostable
potentials, and a general expression to calculate the nonlinear
relaxation time in barrier crossing problems are derived. Finally we
discuss results on the same characteristics and barrier crossing
problems with L\'{e}vy flights, recently obtained with different
approaches.
\end{abstract}
\date{\today}
 \maketitle

\section{Introduction}

Two kinds of motions can easily be observed in Nature: smooth,
regular motion, like Newtonian motion of planets, and random, highly
irregular motion, like Brownian motion of small specks of dust in
the air. The first kind of motion can be predicted and consequently
can be described in the frame of deterministic
approach. The second one demands the statistical approach.\\
\indent The first man who noted the Brownian motion was the Dutch
physician, Jan Ingen--Housz in 1794, who, while in the Austrian
court of Empress Maria Theresa, observed that finely powdered
charcoal floating on an alcohol surface executed a highly erratic
random motion. A similar observation was made by the Scottish
botanist Robert Brown~[Brown, 1828]. He observed under a microscope
the continuous irregular motion of small particles (sized in some
micrometers and less). The particles moved by disordered
trajectories, their motion did not weaken, did not depend on
chemical properties of a medium, strengthened with increasing medium
temperature, with a diminution of its viscosity and sizes of
particles. But R. Brown considered the motion of the particles (not
being atoms of course) as their own property and said nothing
about atoms or molecules.\\
\indent It should pass almost 8 decades before two physicists Albert
Einstein~[Einstein, 1905] and Marian von Smoluchowski
~[Smoluchovski, 1906] found the physical explanation of Brownian
motion. It was based on consideration of thermal motion of molecules
surrounding the Brownian particle. The history of the further study
of Brownian motion is associated with names of Langevin~[Langevin,
1908], Perrin~[Perrin, 1908], Fokker~[Fokker, 1914], Planck~[Planck,
1917], Uhlenbeck, Ornstein~[Uhlenbeck \& Ornstein, 1930],
Chandrasekhar~[Chandrasekhar, 1943] and other well-known physicists.
However, for the first time the diffusion equation appeared in the
thesis of Louis Bachelier~[Bachelier, 1900], a student of A.
Poincar\'{e}. His thesis, entitled "\emph{The theory of
speculations}", was devoted to the study of
random processes in market prices evolution.\\
\indent It is astonishing, how the same diffusion equation can
describe the behavior of neutrons in a nuclear reactor, the light in
atmosphere, the stock market values rate on financial exchange,
particles of flower dust suspended in a fluid and so on. The fact
that completely different by nature phenomena are described by
identical equations is a direct indication that the matter concerns
not the concrete mechanism of the phenomenon, but rather the same
common quality of whole class {\it of similar phenomena}. The
statement of this quality in terms of physical laws and mathematical
postulates or definitions allows to liberate a given pattern from
details, which are not influencing essentially the physical process,
and to explore the obtained model through general laws. This is a
typical situation for statistical physics and applied mathematics.
The new approaches proposed by Einstein, Smoluchowski and Langevin
to describe the Brownian motion, in fact, open the door to model a
great variety of natural phenomena. At the same time for
mathematicians, whose achievements built {\it the theory of random
processes}, the first object of its application became the Brownian
motion. The major contribution to the mathematical theory of
Brownian motion has been brought by N. Wiener~[Wiener, 1930], who
has proved that the trajectories of Brownian process almost
everywhere are continuous but are not differentiable anywhere. Along
with Wiener the mathematical aspects of Brownian motion were treated
by Markov, Doob, Ka\c{c}, Feller, Bernstein, L\'{e}vy, Kolmogorov,
Stratonovich, It\^{o} and others~[Doob, 1953; Ka\c{c}, 1957; Feller,
1971; L\'{e}vy, 1925, 1965; Kolmogorov, 1941; Stratonovich, 1963,
1967, 1992;
It\^{o}, 1944, 1946, 1965].\\
\indent Two important properties are intrinsic to the homogeneous Brownian
motion: the diffusion packet initially concentrated at a point takes
later the Gaussian form, whose width grows in time as
$t^{1/2}$. This kind of diffusion was called the \textit{normal
diffusion}.\\
\indent Twenty years later Einstein, Smoluchowski and Langevin
works, L. Richardson published the article~[Richardson, 1926] where
he presented empirical data being in contradiction with the normal
diffusion: the size $\Delta$ of an admixture cloud in a turbulent
atmosphere grows in time proportionally to $t^{3/2}$, that is much
faster then in the normal case $(t^{1/2})$. This turbulent diffusion
was the first example of superdiffusion processes, when
$\Delta\propto t^{\gamma}$ with
$\gamma>1/2$.\\
\indent The phenomenon has been interpreted as a diffusion process
with a variable diffusivity $D(r)\propto r^{4/3}$. This Richardson's
"law of four thirds" was grounded theoretically by Russian
mathematicians A. N. Kolmogorov~[Kolmogorov, 1941] and A. M.
Obukhov~[Obukhov, 1941] as a consequence of the self-similarity
hypothesis of locally isotropic turbulence. However, the fact that
the diffusivity should depend not on the coordinates (turbulent
medium is supposed to be homogeneous in average), but on the scale
or distance $l$ between a pair of diffusing particles, creates
essential difficulties both to find a solution
to the equation and for its interpretation.\\
\indent As Monin showed in Ref.~[Monin, 1955], the same law of the
diffusive packet widening with time can be obtained in the framework
of the homogeneous Markovian processes family, when the
characteristic function of the spatial distribution of the diffusive
substance

\begin{equation}
\widetilde{P}(k,t) = \left\langle e^{ikX(t)}\right\rangle =
\int_{-\infty}^{+\infty} e^{ikx}P(x, t)dx
\end{equation}

\noindent obeys the equation

\begin{equation}
\frac {\partial \widetilde{P}(k, t)}{\partial t} = - C|k|^\alpha
\widetilde{P}(k, t) \label{diff eq-char fun 1}
\end{equation}

\noindent with $C$ being a positive constant. Under initial
condition $P(x,0)=\delta(x)$ we obtain from Eq.~(\ref{diff eq-char
fun 1})
\begin{equation}
\widetilde{P}(k, t)=\exp \{- C|k|^\alpha t\},
\end{equation}
where $\alpha=2/3$. This is nothing but the characteristic function
of the $\alpha$-stable L\'{e}vy distribution, and the random process
itself is the L\'{e}vy motion (L\'{e}vy flights). Later, anomalous
diffusion in the form of L\'{e}vy flights has been discovered in
many other physical, chemical, biological, and financial
systems~[Shlesinger \emph{et al.}, 1993; Metzler \& Klafter, 2000,
2004; Metzler \emph{et al.}, 2007; Brockmann \& Sokolov, 2002;
Eliazar \& Klafter, 2003; Barkai, 2004;
Chechkin \emph{et al.}, 2006; Mandelbrot, 1997; Mantegna, 1991].\\
\indent L\'{e}vy flights are stochastic processes characterized by
the occurrence of extremely long jumps, so that their trajectories
are not continuous anymore. The length of these jumps is distributed
according to a L\'{e}vy stable statistics with a power law tail and
divergence of the second moment. This peculiar property strongly
contradicts the ordinary Brownian motion, for which all the moments
of the particle coordinate are finite. The presence of anomalous
diffusion can be explained as a deviation of the real statistics of
fluctuations from the Gaussian law, giving rise to the
generalization of the central limit theorem by L\'{e}vy and
Gnedenko~[L\'{e}vy, 1925, 1937; Gnedenko \& Kolmogorov, 1954]. The
divergence of the L\'{e}vy flights variance poses some problems as
regards to the physical meaning of these processes. However,
recently the relevance of the L\'{e}vy motion appeared in many
physical, natural and social complex systems. The L\'{e}vy type
statistics, in fact, is observed in various scientific areas, where
scale invariance phenomena take place or can be suspected. Among
many interesting examples we cite here chaotic dynamics of complex
systems~[Zaslavsky, 2005; Solomon \emph{et al.}, 1993, 1994],
diffusion and annihilation reactions of L\'{e}vy flights with
bounded long-range hoppings~[Albano, 1991], front dynamics in
reaction-diffusion systems with L\'{e}vy
flights~[del-Castillo-Negrete \emph{et al.}, 2003], fractional
diffusion ~[West \emph{et al.}, 1997; Chaves, 1998], thermodynamics
of anomalous diffusion~[Zanette \& Alemany, 1995], dynamical
foundation on noncanonical equilibrium~[Annunziato \emph{et al.},
2001], quantum fractional kinetics~[Kusnezov \emph{et al.}, 1999],
trapping diffusion~[V\'{a}zquez \emph{et al.}, 1999] , L\'{e}vy
diffusion processes as a macroscopic manifestation of
randomness~[Grigolini \emph{et al.}, 1999; Bologna \emph{et al.},
1999], diffusion by flows in porous media~[Painter, 1996],
two-dimensional L\'{e}vy flights~[Desbois, 1992], persistent
L\'{e}vy motion~[Chechkin \& Gonchar, 2000], self-avoiding L\'{e}vy
flights~[Grassberger, 1985], L\'{e}vy flights with quenched noise
amplitudes~[Kutner \& Maass, 1998], cooling down L\'{e}vy
flights~[Pavlyukevich, 2007], branching annihilating L\'{e}vy
flights~[Albano, 1996], random L\'{e}vy flights in the kinetic Ising
and spherical models~[Bergersen \& R\'{a}cz, 1991; Xu \emph{et al.},
1993], plane rotator in presence of a L\'{e}vy random
torque~[C\'{a}ceres, 1999], fluctuations and transport in
plasmas~[Chechkin \emph{et al.}, 2002b; Lynch \emph{et al.}, 2003],
transport in stochastic magnetic fields~[Zimbardo \& Veltri, 1995],
L\'{e}vy flights in the Landau-Teller model of molecular
collisions~[Carati \emph{et al.}, 2003], subrecoil laser
cooling~[Bardou \emph{et al.}, 1994, 2002; Reichel \emph{et al.},
1995; Schaufler \emph{et al.}, 1997, 1999], scintillations and
L\'{e}vy flights through the interstellar medium~[Boldyrev \& Gwinn,
2003], L\'{e}vy flights in cosmic rays~[Wilk \& Wlodarczyk, 1999],
anomalous diffusion in the stratosphere~[Seo \& Bowman, 2000], long
paleoclimatic time series of the Greenland ice core
measurements~[Ditlevsen, 1999a], seismic series and
earthquakes~[Posadas \emph{et al.}, 2002; Sotolongo-Costa \emph{et
al.}, 2000], signal processing~[Nikias \& Shao, 1995], time series
statistical analysis of DNA~[Scafetta \emph{et al.}, 2002], primary
sequences of proteinlike copolymers~[Govorun \emph{et al.}, 2001],
spatial gazing patterns of bacteria~[Levandowsky \emph{et al.},
1997], L\'{e}vy-flight spreading of epidemic processes~[Janssen
\emph{et al.}, 1999], contaminant migration by bioturbation~[Reible
\& Mohanty, 2002], flights of an albatross~[Viswanathan \emph{et
al.}, 1996; Edwards \emph{et al.}, 2007], fractal time in animal
behaviour of \emph{Drosophila} and animal locomotion~[Cole, 1995,
Seuront \emph{et al.}, 2007], financial time series~[Mandelbrot,
1963; Bouchaud \& Sornette, 1994; Mantegna \& Stanley, 1996, 1998;
Chowdhury \& Stauffer, 1999], human stick balancing and L\'{e}vy
flights~[Cabrera \& Milton, 2004] and human memory retrieval as
L\'{e}vy foraging~[Rhodes \& Turvey, 2007]. Experimental evidence of
L\'{e}vy processes was also observed in the motion of single ion in
one-dimensional optical lattice~[Katori \emph{et al.}, 1997] and in
the particle evolution along polymer chains~[Sokolov \emph{et al.},
1997; Lomholt \emph{et al.}, 2005], and in self--diffusion in
systems of polymerlike breakable micelles [Ott \emph{et al.}, 1990].\\
\indent L\'{e}vy flights are a special class of Markovian processes,
therefore the powerful methods of the Markovian analysis are in
force in this case. We mean a possibility to investigate the
stationary probability distributions of superdiffusion, the first
passage time and the residence time characteristics, the spectral
characteristics of stationary motion, etc. Of course, this type of
diffusion has a lot of peculiarities different from those observed
in normal Brownian motion. The main difference from ordinary
diffusion consists in replacing the white Gaussian noise source in
the underlying
Langevin equation with a L\'{e}vy stable noise.\\
\indent In the first part of the present paper, we give a short
introduction to the L\'{e}vy motion. Being a generalization of the
Brownian diffusion, it takes an intermediate place between Brownian
motion and L\'{e}vy processes (i.e. infinitely divisible processes,
see~[Bertoin, 1996]) in the random process hierarchy system. The
L\'{e}vy motion is introduced as a self-similar
L\'{e}vy process.\\
\indent The second part of this paper is devoted to the stationary
probabilistic characteristics and the problem of barrier crossing
for L\'{e}vy flights. We use functional approach to derive the
generalized Kolmogorov equation directly from Langevin equation with
the L\'{e}vy process~[Dubkov \& Spagnolo, 2005]. In particular case
of L\'{e}vy stable noise source we obtain the Fokker-Planck equation
with fractional space derivative. Starting from this equation we
find the exact stationary probability distribution (SPD) of fast
diffusion in symmetric smooth monostable potentials for the case of
Cauchy stable noise. Specifically, we consider potential profiles
$U(x) = \gamma x^{2m}/(2m)$, with odd and even $m$, useful to
describe the dynamics of overdamped anharmonic oscillator driven by
L\'{e}vy noise. We find that for L\'{e}vy flights in steep potential
well, with potential exponent $2m$ greater or equal to four, the
variance of the particle coordinate is finite~[Dubkov \& Spagnolo,
2007]. This gives rise to a confined superdiffused motion,
characterized by a bimodal stationary probability density, as
previously reported in Refs.~[Chechkin \emph{et al.}, 2002a, 2003a,
2004, 2006]. Here we analyze the SPD as a function of a
dimensionless parameter $\beta$, which is the ratio between the
noise intensity $D$ and the steepness $\gamma$ of the potential
profile. We find that the SPDs remain bimodal with increasing
$\beta$ parameter, that is with decreasing the steepness $\gamma$ of
the potential profile,
or by increasing the noise intensity $D$.\\
\indent The particle escape from a metastable state, and the first
passage time density have been recently analyzed for L\'{e}vy
flights in Refs.~[Ditlevsen, 1999b; Rangarajan \& Ding, 2000a,
2000b; Buldyrev \emph{et al.}, 2001; Chechkin \emph{et al.}, 2003b,
2005, 2006, 2007; Dybiec \& Gudowska-Nowak, 2004; Dybiec \emph{et
al.}, 2006, 2007; Bao \emph{et al.}, 2005; Ferraro \& Zaninetti,
2006; Imkeller \& Pavlyukevich, 2006; Imkeller \emph{et al.}, 2007;
Koren \emph{et al.}, 2007]. The main focus in these papers is to
understand how the barrier crossing behavior, according to the
Kramers law [Kramers, 1940], is modified by the presence of the
L\'{e}vy noise. Finally we discuss briefly some results on the
barrier crossing events with L\'{e}vy flights, recently obtained
with different approaches.

\section{L\'{e}vy processes}

To see better a place of the diffusion processes under
consideration among other random processes we shall remind some
definitions. We restrict ourselves to the one-dimensional case
when $X,x\in (-\infty,\infty) $ and $t\ge 0$.\\
\indent A {\it random process} $ \left \{{X(t),t \ge 0} \right \} $
is a set of random variables $X\left ({t} \right) $, given on the
same probability space and corresponding
to any possible time $t\ge 0$.\\
\indent A random process {\it} $ \left \{{X(t),t\ge 0} \right \} $
is called a {\it Markovian process}, if for any $n \ge 1$ and
$t_1<t_2<\dots<t_n<t $ ${\textsf P}(X(t)<x\ |X(t _ 1)=x_1,\dots,
X(t_n)=x_n) = {\textsf P}(X(t)<x|X(t_n)=x_n)$. The Markovian
property is interpreted as independence of future from the past for
the known present. P. L\'{e}vy states this property by the sentence
"the past influences the future only through the present" and
underlines analogy to Huygens' principle ("it is possible to say,
that it is {\it Huygens' principle in calculation
of probabilities}"~[L\'{e}vy, 1965]).\\
\indent A random process {\it} $\left \{X(t),t \in T \right\} $ is
called {\it the process with independent increments} if for any $n
\ge 1$ and $t _ {1} < t _ {2} <\dots < t _ {n} < t $ random
variables $X\left ({0} \right), \; X\left ({t _ {1}} \right) -
X\left ({0} \right),\dots, X\left ({t _ {n}} \right) - X\left ({t _
{n - 1}} \right) $ are mutually independent. The random variable
$X\left ({0} \right) $ is called {\it the initial state (value)} of
the process, and its probability distribution is called {\it initial
distribution} of the process. P. L\'{e}vy named such processes {\it
additive}.
Obviously, they belong to the class of Markovian processes.\\
\indent A random process with independent increments is called {\it
homogeneous} or \textit{stationary}, if the random variables $X\left
({t + \tau} \right) - X\left ({t} \right) $ have distributions which
are independent on $t$:
\begin{equation}
\textsf{P} \left \{{X\left ({t + \tau} \right) - X\left ({t} \right)
< x} \right \} = F\left ({x,\tau} \right).
\end{equation}
P. L\'{e}vy named such processes {\it linear}, remarking that
among them "\emph{there are also processes distinct from
Brownian}". Now, Bertoin~[Bertoin, 1996] and Sato~[Sato, 1999] use
the term \textit{L\'{e}vy processes} for the processes with
stationary independent increments.\\
\indent One can paraphrase the definition by saying that $\{X(t),
t\geq 0\}$ is a L\'{e}vy process if, for every $t,\tau\geq 0$, the
increment $X(t+\tau)-X(t)$ is independent on the process $\{X(t'),
0\leq t'< t\}$ and has the same law as $X(\tau)$. In particular,
$X(0)=0$.\\
\indent We will denote the L\'{e}vy process $L(t)$. As it follows
from the evident decomposition

\begin{eqnarray}
L(t)&=&L \left(\frac{t}{n}\right)+\left[L\left(\frac{2t}{n}\right)
-L\left(\frac{t}{n}\right)\right] \\
&+&\dots
+\left[L\left(\frac{nt}{n}\right)-L\left(\frac{(n-1)t}{n}\right)\right],
\nonumber
\end{eqnarray}

\noindent the random variable $L(t)$ can be divided into the sum of
an arbitrary number of independent and identically distributed
random variables. In other words, the probability distribution of
$L(t)$ belongs to the class of infinitely divisible
distributions~[de Finetti, 1929, 1975; Khintchine, 1938; Khintchine
\& L\'{e}vy, 1936; L\'{e}vy, 1937; Gnedenko \& Kolmogorov, 1954;
Feller, 1971; Mainardi \& Rogosin, 2006]. Hence, we can express the
\textit{second characteristics}, i.e. the logarithm of
characteristic function of the random variable $L(t)$ in the
L\'{e}vy--Khinchine form [Feller, 1971]

\begin{eqnarray}
\phi\left(k,t\right)\equiv \ln\widetilde{P}(k,t) =\ln\left\langle
e^{ikL(t)}\right\rangle &\nonumber \\
=\int_{-\infty}^{+\infty} \left(e^{ikx}-1-ik\sin
x\right)\frac{\rho(x,t)}{x^2}dx , \label{Levy}
\end{eqnarray}

\noindent where $\rho(x,t)$ is the canonical measure density (with
respect
to the first argument).\\
\indent For two consecutive non-overlapping time intervals $t_1$
and $t_2$ we have
\begin{equation}
L(t_1+t_2)\stackrel{d}{=}L(t_1)+L(t_2),
\end{equation}
where $L(t_1)$ and $L(t_2)$ are mutually independent random
variables and the symbol $ \mathop {=} \limits ^{d} $ means the
equality of distributions of the corresponding random variables.
Therefore,
\begin{equation}
\widetilde{P}\left(k,t_1+t_2\right)=
\widetilde{P}(k,t_1)\widetilde{P}(k,t_2) \label{ppp}
\end{equation}
or
\begin{equation}
\phi(k,t_1+t_2) =\phi(k,t_1) +\phi(k,t_2) .
\label{fert}
\end{equation}
According to Eqs.~(\ref{Levy}) and (\ref{fert}) we have
\begin{equation}
\rho(x,t_1+t_2)=\rho(x,t_1)+\rho( x,t_2).\label{Gamel}
\end{equation}
The differentiable solution of Eq.~(\ref{Gamel}), regarding $t$,
is only linear one
\begin{equation}
\rho\left(x,t\right)=t\rho\left(x\right) .
\end{equation}
So, from Eq.~(\ref{Levy}) we obtain
\begin{equation}
\phi\left( k,t\right)
=t\int_{-\infty}^{+\infty}\left(e^{ikx}-1-ik\sin x\right)
\frac{\rho(x)}{x^2}dx,\label{Ch-F}
\end{equation}
where the kernel $\rho\left( x\right) \geq0$. Note that the last
term in the bracket, $-ik\sin x$, serves to ensure the convergence
of the integral and can be omitted if the integral converges
itself. Choosing
\begin{equation}
\rho(x)=\delta(x),
\end{equation}
and taking into account that
\begin{equation}
e^{ikx}-1-ik\sin x=-k^2 x^2/2+o(x^2),\ x\to 0,
\end{equation}
we arrive at the normalized Brownian motion $B(t)$ (Wiener
process) with characteristic function
\begin{equation}
\widetilde{P}(k,t)=\exp\{-tk^2/2\}.
\end{equation}

\section{Self-similarity (scaling)}

The {\it self-similarity} ({\it scaling} is a synonym of
self-similarity) is a special symmetry of a system (process)
revealing that the modification of the scales of one variable can be
compensated by the homothetic transformation of the others. For
example, if the state of a system is characterized by function
$u\left ({x, t} \right) $, where $x$ is the coordinate, $t $ is the
time, the requirement of invariance with respect to scale
transformations $x \to x ' = kx$ and $t \to t ' = lt $, looks like
\begin{equation}
u\left ({x, t} \right) = k ^ {\alpha} l ^ {\delta} u\left ({kx, lt}
\right),
\end{equation}
where $k $ and $l $ are positive, and $ \alpha $ and $ \delta $ are
arbitrary numbers. By choosing $k^{\alpha} = l = m/t$, where $m > 0
$ is a parameter of similarity, we obtain a self-similar form for
the function $u(x,t)$
\begin{equation}
u\left ({x, t} \right) = \left ({{{m} \mathord {\left/ {\vphantom
{{m} {t}}} \right. \kern-\nulldelimiterspace} {t}}} \right) ^ {1 +
\delta} u\left ({\left ( { {{m} \mathord {\left/ {\vphantom {{m}
{t}}} \right. \kern-\nulldelimiterspace} {t}}} \right) ^ {1/\alpha}
x, m} \right). \label{self sim princ}
\end{equation}
In our case such a function is the probability density function
$P(x, t)$. The normalization condition

\begin{equation}
\int_{-\infty}^{+\infty} {P\left ({x, t} \right) dx = 1}
\end{equation}

\noindent and the principle of self-similarity (\ref{self sim
princ}) give $1 + \delta = 1/\alpha$ and lead to the representation
($m=1$)
\begin{equation}
P\left ({x, t} \right) = t ^ {-1/\alpha} g^{(\alpha)}\left ({xt ^
{-1/\alpha}} \right),
\end{equation}
where
\begin{equation}
g^{(\alpha)}\left ({x} \right) = P\left ({x, 1} \right).
\end{equation}
It must be emphasized that we can define the self--similarity of a
random process $X(t)$ with stationary increments as usually (see,
for example, Ref.~[Chechkin et al., 2002c])
$$
X\left( t+\kappa \tau \right) -X\left( t\right) \mathop {=}
\limits^{d} \kappa^H\left[ X\left( t+\tau \right) -X\left(
t\right)\right].
$$
\noindent In such a case $H=1/\alpha $. In terms of characteristic
functions we have
\begin{equation}
\tilde {P} \left ({k, t} \right) = \tilde {g}^{\left ({\alpha}
\right)} \left ( { kt ^ {1/\alpha}} \right),
\end{equation}
with $\tilde{g}^{\left ({\alpha} \right)}$ obeying the equation
\begin{equation}
\tilde{g}^{\left ({\alpha} \right)} \left ({k\left ({t _ {1} + t _
{2}} \right) ^ {1/\alpha}} \right) = \tilde {g} ^ {\left ({\alpha}
\right)} \left ( { kt _ {1} ^ {1/\alpha}} \right) \tilde {g} ^
{\left ({\alpha} \right)} \left ( { kt _ {2} ^ {1/\alpha}} \right)
\label{stability}
\end{equation}
which follows from Eq.~(\ref {ppp}).\\
\indent Let $Y ^ {\left ({\alpha} \right)} $ be a random variable
described by the characteristic function

\begin{equation}
\tilde {g}^{\left ({\alpha} \right)} \left ({k} \right) =
\left\langle \exp\left \{ikY^{(\alpha)}\right\}\right\rangle.
\label{Cha-F 1}
\end{equation}

\noindent Obviously,

\begin{equation}
\tilde {g} ^ {\left ({\alpha} \right)} \left ({kt ^ {1/\alpha}}
\right) = \left\langle\exp\left \{{ikY ^ {\left ({\alpha} \right)}
t ^ {1/\alpha}} \right \}\right\rangle
\end{equation}

\noindent determines the random variable $t ^ {1/\alpha} Y ^ {\left
({\alpha} \right)} $ satisfying the relation

\begin{equation}
\left ({t _ {1} + t _ {2}} \right) ^ {1/\alpha} Y ^ {\left
({\alpha} \right)} \mathop {=} \limits ^ {d} t _ {1} ^ {1/\alpha}
Y _ {1} ^ {\left ({\alpha} \right)} + t _ {2} ^ {1/\alpha} Y _ {2}
^ {\left ({\alpha} \right)},
\end{equation}

\noindent where $Y _ {1} ^ {\left ({\alpha} \right)} $ and $Y _ {2}
^ {\left ({\alpha} \right)} $ are independent copies of random
variable $Y ^ {\left ({\alpha} \right)} $. This relation is a definition
property of {\it strictly stable random variables} with a
characteristic index $ \alpha $. We arrive, therefore, at the very
important subfamily of the L\'{e}vy processes called the
\textit{L\'{e}vy motion} (often, the term "L\'{e}vy flights" is used
as a synonym).

\section{Stable random variables}

To find an explicit expression for the stable characteristic
functions, we can proceed by two equivalent ways: (i) using the
general representation of infinitely divisible characteristic
functions (\ref{Ch-F}) or (ii) using the stability property
(\ref{stability}). We choose the
latter way.\\
\indent Let us introduce the second characteristic
\begin{equation}
\psi ^ {\left ({\alpha} \right)} \left ({k} \right) = \ln\tilde {g}
^ {\left ( {\alpha} \right)} \left ({k} \right),
\label{II Char-F}
\end{equation}
so the property (\ref{stability}) of a strict stability takes the
form
\begin{equation}
\psi ^ {\left ({\alpha} \right)} \left ({\lambda _ {1} k} \right) +
\psi ^ {\left ({\alpha} \right)} \left ({\lambda _ {2} k} \right) =
\psi ^ {\left ( {\alpha} \right)} \left ({\lambda k} \right),
\end{equation}
where
\begin{equation}
\lambda = \left ({\lambda _ {1} ^ {\alpha} + \lambda _ {2} ^
{\alpha}} \right) ^ {1/\alpha}.
\end{equation}
Extending this relation to the sum of arbitrary number $n $ of
identically distributed ($ \lambda _ {1} = \lambda _ {2} = \dots =
\lambda _ {n} = 1 $) terms, we obtain
\begin{equation}
n\psi ^ {\left ({\alpha} \right)} \left ({k} \right) = \psi ^ {\left
({\alpha } \right)} \left ({n ^ {1/\alpha} k} \right).
\end{equation}
According to the property
\begin{equation}
\psi ^ {\left ({\alpha} \right)} \left ({- k} \right) = \left [{\psi
^ {\left ({\alpha} \right)} \left ({k} \right)} \right] ^ {\ast}
\end{equation}
it is enough to determine the function $ \psi ^ {\left ({\alpha}
\right)} \left ({k} \right) $ for positive arguments $k>0$. Taking
into account its continuity in a neighborhood of the origin and the
initial condition of the characteristic function
\begin{equation}
\psi ^ {\left ({\alpha} \right)} \left ({0} \right) = 0,
\end{equation}
we obtain that
\begin{equation}
\left | {\psi ^ {\left ({\alpha} \right)} \left ({k} \right)} \right
| = {\rm const} \cdot k ^ {\alpha} \quad \left ({k > 0, \; \alpha >
0} \right)
\end{equation}
and
\begin{equation}
\psi ^ {\left ({\alpha} \right)} \left ({k} \right) = - k ^ {\alpha}
\left (c _0 - ic_1 \right),
\label{II Char-F b}
\end{equation}
where $c_0$ and $c_1$ are arbitrary real constants. Since the
characteristic function satisfies the requirement
\begin{equation}
\left | {\tilde {g}^{(\alpha)} (k)} \right | \le 1,
\end{equation}
then
\begin{equation}
{ \rm Re} \left\{ \psi^{(\alpha)} (k) \right \} \le 0
\end{equation}
and the real constant $c_{0}$ should be positive. On the other hand,
from Eqs.~(\ref{Cha-F 1}) and~(\ref{II Char-F}) we have
\begin{equation}
\tilde{g}^{(\alpha)}{'}(0) = i\langle Y^{(\alpha)} \rangle, \;\;\;
\tilde{g}^{(\alpha)}{''}(0) = - \langle [Y^{(\alpha)}]^2 \rangle
\vspace{0.001cm}
\end{equation}
and
\begin{eqnarray}
\psi^{(\alpha)}{''}(0) &=& - \langle [Y^{(\alpha)}]^2\rangle +
\langle Y^{(\alpha)}\rangle^2 \nonumber \\
&\equiv& - \textsf{Var}\left\{ Y^{(\alpha)}\right\} \le 0.
\end{eqnarray}
Calculating the second derivative from Eq.~(\ref{II Char-F b}), we
obtain
\begin{equation}
\psi^{(\alpha)}{''}(k) = - (c_0 - ic_1)\alpha(\alpha - 1)k^{\alpha -
2}.
\label{sec der}
\end{equation}
As one can see from Eq.~(\ref{sec der}), we have: (i) for $ \alpha =
2 $ the variance is finite (thus the constant $c _ {1} $ should be
equal to zero since the variance is real); (ii) for $ \alpha < 2 $
and $k \to 0$ we obtain the infinite variance (in this case $c _ {1}
$ does not play any role), and (iii) for $ \alpha > 2 $ and $k \to 0
$ the derivative gives zero. This means that in the expression for
the second moment
$$
\langle [x^{(\alpha)}]^2\rangle = \int_{-\infty}^{+\infty} x^{2}
g^{(\alpha)}(x) dx,
$$
the function $g^{(\alpha)}(x)$ ceases to be a probability density,
when the characteristic index exceeds the boundary value $ \alpha =
2$. We come to the conclusion that a range of values of the
characteristic index $ \alpha $ is the interval $ \left ({0,2}
\right] $ closed on the right. Because of $c_0 > 0$ and $-\infty <
c_1 < +\infty$, the constants of Eq.~(\ref{II Char-F b}) can be put
in the form
\begin{equation}
c _ {0} = 1, \;\; \ c _ {1} = \beta \; {\rm tg} \ \left ({\alpha \pi
/2} \right), \quad - 1 \le \beta \le 1.
\end{equation}
Therefore, the characteristic function $\tilde {g}^{(\alpha,
\beta)}(k)$ of the strictly stable probability distribution
$g^{(\alpha,\beta)}(x)$, with parameters $ \alpha $ and $ \beta $,
is given by the formula

\begin{widetext}
\begin{equation}
\tilde {g}^{(\alpha, \beta)}(k) = \exp\left \{ - |{k}|^{\alpha}
\left [1 - i\beta \; \tan \left (\frac{\alpha \pi}{2} \right) {\rm
sgn}~k \right] \right \}, \label{Stable_Char-F}
\end{equation}
\end{widetext}

\noindent where sgn $x$ is the sign function. The characteristic
index $ \alpha $ (with $ \alpha < 2 $) determines the decreasing
rate of the large values probability for stable distributions
\begin{equation}
\textsf{P} \left \{ \left | Y ^ {(\alpha, \beta)} \right | \ge
\Delta \right\} \propto \Delta^{-\alpha}, \quad \quad \Delta \to
\infty. \label{decr rate}
\end{equation}
The parameter $ \beta $ characterizes the asymmetry of the
distributions: for $ \beta = 0 $ the stable distribution is
symmetric. The class of the symmetric stable distributions, besides
the above-mentioned Gaussian distribution, includes also the Cauchy
distribution
\begin{equation}
g ^ {\left ({1,0} \right)} \left ({x} \right) = \frac {1} {\pi \left
(1 + x ^ {2} \right)}
\label{Cauchy}
\end{equation}
with the characteristic function
\begin{equation}
\tilde {g} ^ {\left ({1,0} \right)} \left ({k} \right) = \exp\left
\{- \left | k \right | \right \}. \label{Cauchy_Cha-F}
\end{equation}

For $ \alpha < 1 $ the distributions with extreme values of
asymmetry $\beta$ are located on a semi-axes only: positive if
${\beta = 1}$ or negative if ${\beta = - 1}$. One of these
well-known one-side distribution is the L\'{e}vy - Smirnov
distribution
\begin{equation}
g ^ {\left ({1/2,1} \right)} \left ({x} \right) = \frac {{1}}
{{\sqrt {2\pi} }} x ^ {- 3/2} \exp\left ({- \frac {{1}} {{2x}}}
\right) {\kern 1pt}, \quad x \ge 0. \label{Levy-Smirnov}
\end{equation}

The detailed exposition of properties of stable random variables and
their distributions can be found in the books~[Khintchine, 1938;
L\'{e}vy, 1965; Gnedenko \& Kolmogorov, 1954; Feller, 1971; Bertoin,
1996; Sato, 1999; Uchaikin \& Zolotarev, 1999b]. We shall underline
here only the fact that all members of the set of stable
distributions are characterized by presence of "heavy" (power-type)
tails and, as a consequence, of infinite variance, and that concerns
all of them, except the Gaussian (normal) distribution. From the
point of view of the whole "noble family", the Gaussian distribution
should be looking defiantly abnormal, monstrous, ugly duckling among
white swans. For us (at least, for many of us) the infinite variance
associates with an infinite error (what is not the truth), with an
infinite energy or with something else what does obviously not have
any physical sense.\\
\indent Professional physicists will recognize in Cauchy density a
natural profile of a radiation line or the cross-section formula for
resonances in nuclear reactions, and they will remember the
Holtzmark distribution describing fluctuations of electric field
strength created by Poisson ensemble of point-like ions in plasma
and the fluctuations of the gravitation field of stellar systems.
But at this stage their acquaintance with the stable laws usually
comes to the end. This should be caused by the circumstance that the
stable densities, as a rule, are not expressed in terms of
elementary functions: the above mentioned formulas fully exhaust a
set of "convenient" distributions. There are some more distributions
which are expressed through the known special functions, like the
following one (see Ref.~[Garoni \& Frankel, 2002])

\begin{widetext}
\begin{equation}
g ^ {\left ({2/3,0} \right)} \left ({x} \right) = \frac {{1}}
{{2\sqrt {3\pi} }} \left | {x} \right | ^ {- 1} \exp\left ({\frac
{{2}} {{27}} x ^ {- 2}} \right)  W _ {- 1/2,1/6} \left ({\frac {{4}}
{{27}} x ^ {- 2}} \right) {\kern 1pt},
\end{equation}
\end{widetext}
where $W _ {\mu, \nu} \left ({x} \right) $ is the hypergeometric
Whittaker function. However, in the age of computers the lack of
simple formulas has no so strong importance. In fact, if simple or
complex formulas can be processed by a computer, never mind, it
computes fast and very well, provided that we check the full process.\\
\indent The properties of "anomalous" stable distributions are
really remarkable. If we shall sum up $n$ independent random
variables, distributed under the same stable law, the breadth of
this new distribution will grow proportionally to $n ^{1/\alpha} $,
and the breadth of distribution of arithmetic mean will grow as $n ^
{1/\alpha - 1} $. For $ \alpha = 1 $, when not only the variance
diverges, but also the expectation value does not exist, the width
of arithmetic mean distribution remains constant! And if the results
of your measuring are distributed by the Cauchy law, the repetition
of measuring cannot decrease the "statistical error" in no way. If $
\alpha < 1 $, increasing the number of terms in the sum of results
in widening of "sample mean" distribution! Clearly, the law of large
numbers does not work here because the expectation value does not
exist.

In summary we remark that the whole  set of stable laws appear as
limiting distributions in the {\it generalized central limiting
theorem}: any other laws cannot be limiting ones. This is,
certainly, their  most important advantage.

\section{Stable processes and L\'{e}vy motion}

Having designated the random realization of the processes under
consideration as $L^{(\alpha, \beta)}(t)$, we write the condition of
self-similarity as
\begin{equation}
L^{(\alpha, \beta)}(t)= t^{1/\alpha}Y^{(\alpha, \beta)},
\end{equation}
where $Y^{(\alpha, \beta)} $ is the strictly stable random variable,
with parameters $ \alpha $ and $ \beta $. The set of such processes
is sometimes called {\it stable processes}.

 The random process $ \left \{{X\left ({t} \right), t \in
T} \right \} $ is called {\it stable (strictly stable)}, if all its
finite-dimensional distributions  are stable (strictly stable). This
definition generalizes the concept of Gaussian process, not
restricted by the requirements of homogeneity and independence of
increments. Thus, however, it is necessary to introduce the concept
of multivariate stable distribution or multivariate stable vector.

 The random vector $ {\bf Y} = \left (Y _ {1},\dots, Y _{m}\right)$
is called {\it stable random vector} in $\Re^m$, if for any positive
numbers $ \lambda _ {1}, \lambda _ {2} $ there are positive number $
\lambda $ and vector $ {\bf c} \in \Re^m$, such that
\begin{equation}
\lambda _ {1} {\bf Y} ' + \lambda _ {2} {\bf Y} ''\mathop {=}
\limits ^ {d} \lambda {\bf Y} + \textbf{c},
\end{equation}
where $ {\bf Y}',  {\bf Y}{''} $ are independent copies of the
random vector $ {\bf Y} $.

 The stable random vector $ {\bf Y} $ is called {\it
strictly stable} if the last equality, with $ {\bf c} = 0 $, holds
true for any $ \lambda _ {1} $ and $ \lambda _ {2} $.

The stable random vector $ {\bf Y} $ is called {\it symmetric stable
random vector}, if it satisfies the relation
\begin{equation}
\textsf{P}\{{\bf Y}\in A\} = \textsf{P}\{- {\bf Y} \in A \}
\end{equation}
for any Borel set $A \subset \Re^m $. Similarly to the
one-dimensional case, the symmetric vector is strictly stable (the
inverse statement, certainly, is not true).

Recall the standard definition. The random process $\{
L^{(\alpha,\beta)} (t), t  \ge  0 \}$ is called (standard) $ \alpha
$ {\it -stable Levy-motion with parameters} $ 0 < \alpha \le 2 $, $
- 1 \le \beta \le 1 $, if
\begin{enumerate}
\item $L^{(\alpha,\beta)}\left (0 \right) $ = 0 almost certainly;
\item $ \left \{{L^{(\alpha,\beta)}\left ({t} \right),  t\ge
0} \right \} $ is a process with independent increments;
\item $L^{(\alpha,\beta)}\left ({t + \tau} \right) -
L^{(\alpha,\beta)}\left ({t} \right) \mathop { = } \limits ^ {d}
\tau ^ {1/\alpha} Y ^ {\left ({\alpha, \beta} \right)}\\
$ at any $ t$ and $ \tau $.
\end{enumerate}
For the sake of brevity we shall call it $L ^ {\left ({\alpha,
\beta} \right)} $--{\it process}, then the Wiener process will be
designated as $L ^ {\left ({2,0} \right)} $-process.\\
\indent For better understanding of the principal difference between
$L$-processes with $\alpha = 2 $ and $ \alpha < 2 $, consider the
third property for the Wiener process, namely, the Lindeberg
condition which reflects the continuity of its trajectories.
According to the continuity criterion of a random process
[Lo\`{e}ve, 1963], for $\alpha = 2$ and $\tau\rightarrow 0$ we have
\begin{widetext}
\begin{equation}
\frac{\textsf P \left\{ {\left | {L^{(2,0)}\left ({t + \tau} \right)
- L^{(2,0)}\left ({t} \right)} \right | \ge \Delta} \right\} }{\tau}
= \frac{{\textsf P} \left\{ \left | Y^{\left( 2,0\right)} \right |
\ge \Delta /\sqrt{\tau } \right\} }{\tau} = \frac {{1}} {{\sqrt
{\pi} \tau ^ {}}} \int\limits _ {\Delta /\sqrt {\tau}} ^ {\infty} {e
^ {- z ^ {2} /4} dz}.
\label{Lind-condition}
\end{equation}
\end{widetext}
Evaluating the indeterminate form by l'Hopital's rule, we obtain for
$\tau \to 0$

\begin{equation}
\frac {1} {\sqrt {\pi}} \; \frac {d} {d\tau} \int_{\Delta /\sqrt
{\tau}}^{\infty} {e ^ {- z ^ {2} /4} dz} = \frac {\Delta} {2\sqrt
{\pi}} \tau ^ {- 3/2} e ^ {-\Delta ^ {2} /4\tau} \to 0 .
\end{equation}

\noindent From Eq.~(\ref{decr rate}) and using the property $3$, for
$\alpha < 2$ and $ \tau \to 0 $, we have

\begin{widetext}
\begin{eqnarray}
\frac{{\textsf P} \left\{ \left | L^{(\alpha,\beta)}\left (t + \tau
\right) - L^{(\alpha,\beta)}\left (t \right) \right | \ge \Delta
\right\}}{\tau} &=& \frac{{\textsf P} \left \{\tau^{1/\alpha} \left
| Y^{\left (\alpha, \beta \right)} \right | \ge \Delta \right\}}{\tau} \nonumber \\
&=& \frac{{\textsf P} \left \{\left | Y^{\left (\alpha, \beta
\right)} \right | \ge \Delta \tau^{- 1/\alpha} \right\}}{\tau}
\propto \Delta^{- \alpha} > 0.
\end{eqnarray}
\end{widetext}

Thus, $L ^{(2,0)}$ is the only $L ^{(\alpha,\beta)}$ process
possessing continuous trajectories. As it was shown in
Ref.~[Seshadri \& West, 1982], L\'{e}vy index $\alpha $ is the
fractal dimensionality of the L\'{e}vy process
trajectories.\\
\indent The size of a diffusion package for $ \alpha < 2 $ grows
with time faster than $ \sqrt {t} $, namely,  proportionally to $t ^
{1/\alpha} $, and its shape differs from the Gaussian law. The
variance is infinite, but nothing interferes in using any other
measure of width, for example, the width on the half of peak height,
or the width of the interval containing some fixed probability. For
example, if we consider the fractal moment of L\'{e}vy process
increments
$$
\left\langle \left | L^{(\alpha,\beta)}\left( t\right) -
L^{(\alpha,\beta)}\left( 0\right) \right |^{\delta }\right\rangle
=\int_{-\infty}^{+\infty} P(x,t)|x|^{\delta} dx,
$$
which is finite for $0<\delta <\alpha $, we immediately obtain from
Eq.~(19)
$$
\left\langle \left | L^{(\alpha,\beta)}\left( t\right) -
L^{(\alpha,\beta)}\left( 0\right) \right |^{\delta }\right\rangle
\sim t^{\delta /\alpha },
$$
and, as a result (see Ref.~[Chechkin \emph{et al.}, 2002c])
$$
\left\langle \left | L^{(\alpha,\beta)}\left( t\right) -
L^{(\alpha,\beta)}\left( 0\right) \right |^{\delta }\right\rangle
^{1/\delta }\sim t^{1/\alpha }.
$$
\indent We consider now the stochastic model of {\it L\'{e}vy flight
superdiffusion}.

\section{Fractional equation for L\'{e}vy flight superdiffusion}

If $\alpha=2$, the L\'{e}vy motion becomes the Brownian motion with
characteristic function

\begin{equation}
\widetilde{P}^{(2,0)}(k,t)=e^{-tk^2},
\end{equation}

\noindent obeying the differential equation

\begin{equation}
\frac{\partial \widetilde{P}^{(2,0)}(k,t)}{\partial
t}=-k^2\widetilde{P}^{(2,0)}(k,t),
\end{equation}

\noindent under initial condition
\begin{equation}
\widetilde{P}^{(2,0)}(k,0)=1.
\end{equation}
Factor $-k^2$ is the Fourier image of the one-dimensional Laplace
operator $\triangle_1=\partial^2/\partial x^2$. The inverse
transformation yields the partial differential equation
\begin{equation}
\frac{\partial P^{(2,0)}(x,t)}{\partial t}=\frac{\partial^2
P^{(2,0)}(x,t)}{\partial x^2},
\end{equation}
with initial condition
\begin{equation}
P^{(2,0)}(x,0)=\delta(x).
\end{equation}
For the symmetric L\'{e}vy motion with an arbitrary $\alpha$, the
corresponding expression of the characteristic function reads
\begin{equation}
\widetilde{P}^{(\alpha,0)}(k, t) = e^{-t|k|^\alpha},
\end{equation}
and
\begin{equation}
\frac{\partial \widetilde{P}^{(\alpha,0)}(k, t)}{\partial t}
=-|k|^\alpha\widetilde{P}^{(\alpha,0)}(k, t).
\end{equation}
Taking into account that $-|k|^\alpha$ is the Fourier image of the
Riesz fractional operator
$\triangle_1^{\alpha/2}=\partial^\alpha/\partial |x|^\alpha$, we
arrive at the fractional differential equation
\begin{equation}
\frac{\partial P^{(\alpha,0)}(x,t)}{\partial
t}=\frac{\partial^\alpha P^{(\alpha,0)}(x, t)}{\partial|x|^\alpha}.
\end{equation}
By means of the direct Fourier transformation, one can be convinced
of the validity of two following integral representations of the
Riesz derivative

\begin{widetext}
\begin{equation}
\frac{\partial^\alpha f(x)}{\partial |x|^\alpha} =
-\frac{1}{K(\alpha)}\int_{-\infty}^{+\infty} \frac{f(x) -
f(\xi)}{|x-\xi|^{1+\alpha}} d\xi =
-\frac{1}{K(\alpha)}\int_{0}^{+\infty} \frac{2f(x)
-f(x-\xi)-f(x+\xi)}{\xi^{1+\alpha}}d\xi.
\end{equation}
\end{widetext}

\noindent Here

\begin{equation}
K(\alpha) = \frac{\pi}{\Gamma(\alpha + 1) \sin(\pi\alpha/2)}\,.
\end{equation}

\noindent Finally, in the case of the asymmetric L\'{e}vy motion,
the equation for probability distribution becomes

\begin{equation}
\frac{\partial P^{(\alpha,\beta)}(x,t)}{\partial
t}=D_x^{(\alpha,\beta)} P^{(\alpha,\beta)}(x,t).
\end{equation}

\noindent This equation contains the Feller fractional space
derivative $D_x^{(\alpha,\beta)}$, which is determined by the
relation

\begin{widetext}
\begin{eqnarray}
D_x^{(\alpha,\beta)}f(x)&=& -\frac{A(\alpha,\beta)}{K(\alpha)}
\int_{-\infty}^{+\infty} \frac{1+\beta~{\rm
sgn}~(x-\xi)}{|x-\xi|^{1+\alpha}}[f(x) - f(\xi)]d\xi\nonumber \\
&=& -\frac{A(\alpha,\beta)}{K(\alpha)}\int_{0}^{+\infty} \frac{2f(x)
- (1+\beta)f(x-\xi) - (1 - \beta)f(x+\xi)}{ \xi^{1+\alpha}}d\xi \, ,
\end{eqnarray}
\end{widetext}

\noindent where
\begin{equation}
A(\alpha,\beta) = 1+ \beta^2{\rm tg}(\alpha \pi/2).
\end{equation}
A more detailed consideration of fractional differential equation
for description of L\'{e}vy motion can be found in Refs.~[Saichev \&
Zaslavsky, 1997; Uchaikin, 1999, 2000, 2002, 2003a, 2003b; Uchaikin
\& Zolotarev, 1999; Metzler \& Klafter, 2000; Mainardi \emph{et
al.}, 2001; Metzler \& Nonnenmacher, 2002; Lenzi \emph{et al.},
2003; Gorenflo \& Mainardi, 2005; Sokolov \& Chechkin, 2005;
Zaslavsky, 2005].

\section{L\'{e}vy white noise}

Let us come back to the L\'{e}vy processes. The time derivative of
the L\'{e}vy process

\begin{equation}
\xi\left( t\right) =dL(t)/dt\equiv\dot{L}( t)
\end{equation}
is a stationary random process and has analogy to the Gaussian white
noise, which is the time derivative of the Wiener process. The
L\'{e}vy process, in fact, is a generalized Wiener process.

Now we derive the characteristic functional $\Theta_t[u]$ of this
L\'{e}vy delta-correlated noise. By definition, we have
\begin{widetext}
\begin{eqnarray}
\Theta_t[u]  &=& \left\langle \exp\left\{ i\int_{0}^{t}u\left(
\tau\right) \xi\left(  \tau\right) d\tau\right\} \right\rangle =
\left\langle \exp\left\{ i \int_{0}^{t} u\left(  \tau\right)
dL\left( \tau\right)
\right\}  \right\rangle \nonumber \\
&=& \left\langle \exp\left\{
i\lim\limits_{\delta_{\tau}\rightarrow0}
\sum\limits_{k=1}^{n}u\left(  \vartheta_{k}\right)  \left[ L\left(
\tau_{k}\right)  -L\left(\tau_{k-1}\right) \right] \right\}
\right\rangle \label{Stil}  \\  &=&
\lim\limits_{\delta_{\tau}\rightarrow0}\left\langle
\prod\limits_{k=1} ^{n}\exp\left\{  iu\left(  \vartheta_{k}\right)
\left[ L\left( \tau _{k}\right)  -L\left(  \tau_{k-1}\right) \right]
\right \} \right\rangle = \lim\limits_{\delta_{\tau}\rightarrow0}
\prod\limits_{k=1}^{n}\tilde{P}(u(\vartheta_k),\Delta \tau _k),
\nonumber
\end{eqnarray}
\end{widetext}
where $\vartheta _k$ is some internal point of time interval $\left(
\tau
_{k-1},\tau _k\right) \,$, $\delta _\tau =\max \limits_k\Delta \tau _k$, $%
\Delta \tau _k=\tau _k-\tau _{k-1}$ ($\tau _0=0$, $\tau _n=t$), and
$\tilde{P}(u(\vartheta_k),\Delta \tau _k)$ is the characteristic
function of increments. To obtain Eq.~(\ref{Stil}) we used the
statistical independence of increments of L\'{e}vy process $L \left(
t\right) $. Further from Eqs.~(\ref{Ch-F}) and (\ref{Stil}) we
obtain
\begin{widetext}
\begin{eqnarray}
\Theta_t[u]  &=& \lim\limits_{\delta_{\tau}\rightarrow0}\prod
\limits_{k=1}^{n}\exp\left\{ \Delta\tau_{k}\int_{-\infty}^{+\infty}
\frac{e^{iu\left( \vartheta_{k}\right)  x}-1-iu\left(
\vartheta_{k}\right) \sin
x}{x^{2}}\,\rho\left(  x\right)  dx\right\}  \nonumber \\
&=& \exp\left\{  \lim\limits_{\delta_{\tau}\rightarrow0}\sum\limits_{k=1}%
^{n}\Delta\tau_{k}\int_{-\infty}^{+\infty}\frac{e^{iu\left(
\vartheta_{k}\right)  x}-1-iu\left(  \vartheta_{k}\right)  \sin x}{x^{2}}%
\,\rho\left(  x\right)  dx\right\} \nonumber \\
&=& \exp\left\{  \int_{0}^{t}d\tau
\int_{-\infty}^{+\infty}\frac{e^{iu\left(  \tau\right) x}-1-iu\left(
\tau\right)  \sin x}{x^{2}}\,\rho\left(  x\right) dx\right\} .
\label{main}
\end{eqnarray}
\end{widetext}
Now we are going to derive a useful functional formula for the
L\'{e}vy white noise. The formula to split the correlation between
a Gaussian random vector field $\mathbf{\xi}\left( \mathbf{r},
t\right) $ and its arbitrary functional $R[\mathbf{\xi}]$ was for
the first time obtained by Furutsu~[Furutsu, 1963] and
Novikov~[Novikov, 1965]. For Gaussian random process $\xi \left(
t\right) $ with zero mean it reads
\begin{equation}
\left\langle \xi \left( t\right) R\left[ \xi \right] \right\rangle
=\int K\left( t,\tau \right) \left\langle \frac{\delta R\left[ \xi
\right] }{\delta \xi \left( \tau \right) }\right\rangle d\tau ,
\label{FN}
\end{equation}
where $K\left( t,\tau \right) =\left\langle \xi \left( t\right) \xi
\left( \tau \right) \right\rangle$ is the correlation function of
Gaussian noise $\xi\left( t\right) $. We use the generalization of
Furutsu--Novikov formula (\ref{FN}), obtained by
Klyatskin~[Klyatskin, 1974], for arbitrary functional $R_t[\xi]$ of
a non-Gaussian random process $\xi\left(\tau\right)$, defined on the
observation interval $\tau\in\left( 0,t\right) $,
\begin{equation}
\left\langle \xi \left( t\right) R_t[\xi+z]\right\rangle =\left.
\frac{\dot \Phi _t\left[ u\right] }{iu\left( t\right) } \right|
_{\,u=\frac{\delta }{i\delta z}}\left\langle R_t\left[ \xi
+z\right] \right\rangle . \label{Kly}
\end{equation}
Here $z\left( t\right) $ is arbitrary deterministic function, and
$\Phi _t\left[ u\right] =\ln \Theta _t\left[ u\right] $. From
Eq.~(\ref{main}) we obtain the following expression for
variational operator in Eq.~(\ref{Kly})

\begin{widetext}
\begin{equation}
\frac{\dot{\Phi}_{t}\left[ u\right] }{iu\left( t\right) } =
\int_{-\infty}^{+\infty}\frac{e^{iu\left( t\right) x}-1-iu\left(
t\right) \sin x}{iu\left( t\right) x^{2}}\,\rho\left( x\right) dx =
\int_{-\infty}^{+\infty} \frac{\rho\left( x\right) }{x^{2}}\,dx
\int_{0}^{x}[e^{iu\left( t\right) y}-\cos y]dy.
\end{equation}
\end{widetext}

\noindent Substituting this equation in Eq.~(\ref{Kly}) we arrive at

\begin{widetext}
\begin{equation}
\left\langle \xi\left( t\right) R_{t}\left[
\xi+z\right]\right\rangle =\int_{-\infty}^{+\infty}\frac{\rho\left(
x\right) }{x^{2}}\,\,dx\int_{0}^{x}\left[ \exp\left\{
y\frac{\delta}{\delta z\left( t\right) }\right\} -\cos y\right]
\left\langle R_{t}\left[ \xi+z\right] \right\rangle dy\,.
\label{LPro}
\end{equation}
\end{widetext}

\noindent By inserting the operator of functional differentiation
into the average in Eq.~(\ref{LPro}) and by putting $z = 0$, we get
finally

\begin{widetext}
\begin{equation}
\left\langle \xi\left( t\right) R_{t}\left[ \xi\right] \right\rangle
=\int_{-\infty}^{+\infty}\frac{\rho\left( x\right) }{
x^{2}}\,dx\int_{0}^{x}\left[ \left\langle \exp\left\{
y\frac{\delta}{ \delta\xi\left( t\right) }\right\} R_{t}\left[
\xi\right] \right\rangle -\left\langle R_{t}\left[ \xi\right]
\right\rangle \cos y\right] dy\,. \label{LP}
\end{equation}
\end{widetext}

\section{Derivation of Kolmogorov's equation}

Let us consider now the Langevin equation with the L\'{e}vy white
noise source $\xi (t)$
\begin{equation}
\dot X=f\left( X,t\right) +g\left( X,t\right) \xi \left( t\right)
. \label{Lang}
\end{equation}
By differentiating with respect to time the well-known expression
for probability density of the random process $X(t)$
\begin{equation}
P\left( x,t\right) =\left\langle \delta\left( x-X\left( t\right)
\right) \right\rangle , \label{W}
\end{equation}
and taking into account Eq.~(\ref{Lang}), we obtain
\begin{widetext}
\begin{equation}
\frac{\partial P}{\partial t} = -\frac{\partial}{\partial x}
\,\left( f(x,t) P\right) - \frac{\partial}{\partial x}\,\,g(x,t)
\left\langle \xi \left( t\right) \delta\left( x-X\left( t\right)
\right) \right\rangle . \label{prel}
\end{equation}
\end{widetext}
By using functional differentiation rules and following the same
procedure used in Ref.~[H\"{a}nggi, 1978], from Eq.~(\ref{Lang}) we
get
\begin{widetext}
\begin{equation}
\frac{\delta}{\delta\xi \left( t\right) }\,\,\delta\left( x-X\left(
t\right) \right) =-\frac{\partial}{\partial x}\,\,g\left( x,t\right)
\delta\left( x-X\left( t\right) \right) . \label{equi}
\end{equation}
\end{widetext}
Thus, the variational operator $\delta/\delta\xi \left( t\right) $
with respect to the function $\delta\left( x-X\left( t\right)
\right) $ is equivalent to the ordinary differential operator $
-\partial/\partial x\left( g\left( x,t\right) \,\right) $. Taking
into account Eqs.~(\ref {LP}), (\ref{prel}), and (\ref{equi}), we
obtain, after integration, the following Kolmogorov's equation for
nonlinear system (\ref{Lang}) driven by L\'{e}vy white noise [Dubkov
\& Spagnolo, 2005]
\begin{widetext}
\begin{equation}
\frac{\partial P}{\partial t}=-\frac{\partial \left[ f(x,t)P\right]
}{\partial x}+\int_{-\infty }^{+\infty }\frac{\rho \left( z\right)
}{z^2}\left[ \exp \left\{ -z\frac {\partial }{\partial x}g\left(
x,t\right) \right\} -1+\sin z\frac {\partial }{\partial x}g\left(
x,t\right) \right] dz\,P. \label{gene}
\end{equation}
\end{widetext}

We analyze further some different kernel functions $\rho \left(
x\right) $ to obtain particular cases of Kolmogorov's equation
(\ref{gene}), related to different non-Gaussian white noise sources.\\
\indent (a) As a first simple case we consider a Gaussian white
noise $\xi (t)$. The corresponding kernel function is $\rho \left(
x\right) =2D\delta \left( x\right) $, where $D$ is the noise
intensity. After substituting this kernel in Eq.~(\ref{gene}), we
obtain the ordinary Fokker-Planck equation
\begin{equation}
\frac{\partial P}{\partial t}=-\frac \partial {\partial x}\,\left( f
P\right) + D\frac \partial {\partial x}\,\,g \frac \partial
{\partial x}\,\,\left( g P\right) . \label{FP}
\end{equation}
\\
\indent (b) For additive driving noise $\xi\left( t\right) $,
$g\left( X,t\right) =1$ in Eq.~(\ref{Lang}), and the exponential
operator in Eq.~(\ref{gene}) reduces to the space shift operator.
As a result, we find
\begin{widetext}
\begin{equation}
\frac{\partial P}{\partial t}=-\frac \partial {\partial x}\left[
f\left( x,t\right) P\right] +\int_{-\infty }^{+\infty} \frac{\rho
\left( z\right) }{z^2}\left[ P\left( x-z,t\right) -P\left(
x,t\right) +\sin z \frac{\partial P\left( x,t\right) }{\partial
x}\right] dz. \label{add}
\end{equation}
\end{widetext}
Equation (\ref{add}) is similar to the Kolmogorov-Feller equation
for purely discontinuous Markovian processes [Saichev \& Zaslavsky,
1997; Kami\'{n}ska \& Srokowski, 2004; Dubkov \& Spagnolo, 2005]
\begin{equation}
\frac{\partial P}{\partial t}=\nu \int_{-\infty }^{+\infty} w \left(
x-z\right) P\left( z,t\right) dz-\nu P\left( x,t\right) ,
\label{Kol}
\end{equation}
where $w\left( x\right) $ is the probability density of jumps step,
and $\nu $ is the constant mean rate of jumps. By putting $f\left(
X,t\right) =0$, omitting the term with $sin z$, and comparing
Eq.~(\ref{add}) with Eq.~(\ref{Kol}) we find the kernel function for
such a case
\begin{equation}
\rho \left( x\right) =\nu x^2 w\left( x\right) . \label{conn}
\end{equation}
For non--Gaussian additive driving force $\xi ^{(\alpha)}\left(
t\right) $, with symmetric $\alpha $-stable L\'{e}vy distribution,
the kernel function reads $\rho \left( x\right) = Q \left| x\right|
^{1-\alpha }$. As a result, Eq.~(\ref{add}) takes the following form
\begin{widetext}
\begin{equation}
\frac{\partial P}{\partial t}=-\frac{\partial}{\partial x}%
\,\left[ f\left( x,t\right) P\right] +
Q \int_{-\infty}^{+\infty}\frac{%
P\left( z,t\right) -P\left( x,t\right) }{\left\vert x-z\right\vert
^{\alpha+1}}\,\,dz \label{KE}
\end{equation}
\end{widetext}
and describes the anomalous diffusion in form of symmetric
L\'{e}vy flights.\\
\indent In accordance with the definition of Riesz derivative (60),
Eq.~(82) can be written as
\begin{equation}
\frac{\partial P}{\partial t}=-\frac{\partial}{\partial x}\left[
f\left( x,t\right) P\right]
+D\frac{\partial^{\alpha}P}{\partial\left\vert x\right\vert
^{\alpha}}
\label{FFPE}
\end{equation}
where (see Eq.~(61))
\begin{equation}
D=K\left( \alpha \right) Q=\frac{\pi Q}{\Gamma (\alpha+1)\sin
{(\pi\alpha/2)}}\,.
\end{equation}
\indent For the first time, the fractional Fokker-Planck
equation~(\ref{FFPE}) for L\'{e}vy flights in the potential profile
$U\left( x\right) $ (with $- U^{\prime } = f( x,t)$ was obtained
directly from Langevin equation

\begin{equation}
\dot{X}=-U^{\prime}\left( X\right) +\xi^{(\alpha)}\left( t\right)\,,
\label{LSN}
\end{equation}

\noindent by replacing the white Gaussian noise $\xi\left(
t\right)\equiv\xi^{(2)}\left( t\right) $ with the symmetric L\'{e}vy
$\alpha$-stable noise $\xi^{(\alpha)}\left( t\right) $, in
Refs.~[Ditlevsen, 1999b; Yanovsky \emph{et al.}, 2000; Garbaczewski
\& Olkiewicz, 2000] (see also~[Schertzer \emph{et al.}, 2001]).
However, some attempts were undertaken before in Refs.~[Fogedby,
1994a, 1994b, 1998; Jespersen \emph{et al.}, 1999]. Recently using a
different approach it was derived in [Dubkov \& Spagnolo, 2005].

\section{Stationary probability distributions for L\'{e}vy flights}

First of all, we can try to evaluate the stationary probability
distribution $P_{st}\left( x\right)$ of L\'{e}vy flights in the
potential profile $U(x)$ from Eq.~(\ref{FFPE}). Of course, this
evaluation is impossible for any potential profile, but the
potential $U(x)$ should satisfy some constraints. It is better to
apply Fourier transform to the integro-differential
equation~(\ref{FFPE}) and to write the equation for the
characteristic function

\begin{equation}
\tilde P\left( k,t\right) =\left\langle e^{ikX\left(
t\right)}\right\rangle =\int_{-\infty }^{+\infty}e^{ikx}P\left(
x,t\right) dx. \label{Char-0}
\end{equation}

\noindent After simple manipulations we find (see Eq~(58))

\begin{equation}
\frac{\partial \tilde P}{\partial
t}=-ik\int_{-\infty}^{+\infty}e^{ikx}U^{\prime }(x)P\left(
x,t\right) dx-D\left\vert k\right\vert ^{\alpha }\tilde P
.\,\nonumber
\end{equation}

\noindent  For smooth potential profiles $U\left( x\right)$,
expanding in power series in a neighborhood of the point $x=0$, we
can rewrite Eq.~(87) in the operator form

\begin{equation}
\frac{\partial \tilde P}{\partial t}=-ikU^{\prime }\left(
-i\frac{\partial }{\partial k}\right) \tilde P-D\left\vert
k\right\vert ^{\alpha }\tilde P\,. \label{Char-1}
\end{equation}

\noindent In particular, for stationary characteristic function,
from Eq.~(\ref {Char-1}) we get

\begin{equation}
U^{\prime }\left( -i\frac{d}{dk}\right) \tilde P_{st}-iD\left\vert
k\right\vert ^{\alpha -1}\mathrm{sgn}\,k\cdot \tilde P_{st}=0\,,
\label{St-Char}
\end{equation}

\noindent where $\mathrm{sgn}\,k$ is the sign function.
Unfortunately, one cannot solve analytically Eq.~(\ref{St-Char}) for
arbitrary potential $U\left(
x\right) $ and arbitrary L\'{e}vy exponent $\alpha $.\\
\indent Let us consider, as in Ref.~[Chechkin \emph{et al.}, 2002a],
the symmetric smooth monostable potential $U\left( x\right) =\gamma
x^{2m}/\left( 2m\right) $ $\left( m=1,2,\ldots \right) $. The
Eq.~(\ref{St-Char}), therefore, transforms into the following
differential equation of $\left( 2m-1\right) $-order

\begin{equation}
\tilde P_{st}^{(2m-1)}+\left( -1\right) ^{m+1}\beta
^{2m-1}\left\vert k\right\vert ^{\alpha -1}\mathrm{sgn}\,k\cdot
\tilde P_{st}=0\,, \label{Mon-Pot}
\end{equation}

\noindent where $\beta =\sqrt[2m-1]{D/\gamma }$. As it was proved by
analysis of Eq.~(\ref{Mon-Pot}) for small arguments $k$ in
Ref.~[Chechkin et al., 2002a], the stationary probability
distribution $P_{st}\left( x\right)$ has non-unimodal shape and
power tails

\begin{equation}
P_{st}\left( x\right) \sim \frac{1}{\left\vert x\right\vert
^{2m+\alpha -1}}\,,\qquad \left\vert x\right\vert \rightarrow
\infty \,. \label{Tail}
\end{equation}

\noindent According to Eq.~(\ref{Tail}), we have a confinement of
L\'{e}vy flights (finite variance of particle's coordinate) in the
case when
\begin{equation}
2m>4-\alpha .\label{conf}
\end{equation}
\indent Because of the L\'{e}vy index $\alpha \in (0,2]$, a
confinement takes place for all values of $\alpha $, starting from
quartic potential $(m=2)$. Exact solution of Eq.~(\ref{Mon-Pot}) can
be only obtained for the case of Cauchy noise source $\xi
^{(1)}\left( t\right) $ $(\alpha =1)$. Due to the symmetry of the
characteristic function $\tilde P_{st}\left( -k\right) =\tilde
P_{st}\left( k\right) $, we can reduce Eq.~(\ref{Mon-Pot}) to a
linear differential equation with constant parameters

\begin{equation}
\tilde P_{st}^{(2m-1)}-\left( -1\right) ^{m}\beta ^{2m-1}\tilde
P_{st}=0\qquad \left( k>0\right) . \label{Cauchy-dif.eq.}
\end{equation}

\noindent From the corresponding characteristic equation

\begin{equation}
\lambda ^{2m-1}=\left( -1\right) ^{m}\beta ^{2m-1},
\end{equation}

\noindent we select the roots with negative real part, which are
meaningful from physical point of view. The general solution of
Eq.~(\ref{Cauchy-dif.eq.}), therefore, reads

\begin{widetext}
\begin{equation}
\tilde P_{st}\left( k\right) =\sum_{l=0}^{\left[ \left( m-1\right)
/2 \right] }A_{l}\exp \left\{ -\beta \left\vert k\right\vert \cos
\frac{\pi \left( m-2l-1\right) }{2m-1}\right\}\cdot \nonumber\\
\cos \left( \beta \left\vert k\right\vert \sin \frac{\pi \left(
m-2l-1\right) }{2m-1}-\varphi _{l}\right) ,
\label{General}
\end{equation}
\vspace{0.1cm}
\end{widetext}

\noindent where the quadratic brackets in the upper limit of the sum
$[(m-1)/2]$ denote the integer part of the enclosed expression. The
unknown constants $A_{l}$ and $\varphi _{l}$ can be calculated from
the conditions

\begin{equation}
\tilde P_{st}\left( 0\right) = 1,\quad \tilde P_{st}^{\left(
2j-1\right) }\left( +0\right) = 0 ,
\label{Cond}
\end{equation}

\noindent where $j=1,2,\ldots ,m-1$. Now substituting
Eq.~(\ref{General}) in Eq.~(\ref{Cond}) we obtain

\begin{widetext}
\begin{equation}
\sum_{l=0}^{\left[ \left( m-1\right) /2\right] }A_{l}\cos \varphi
_{l}=1, \; \sum_{l=0}^{\left[ \left( m-1\right) /2\right] }A_{l}\cos
\left[ \frac{\pi \left( 2j-1\right) \left( m+2l\right)
}{2m-1}-\varphi _{l}\right] =0 \quad (j=1,2,\ldots ,m-1) .
\label{Constants}
\end{equation}
\vspace{0.1cm}
\end{widetext}

\noindent Making the reverse Fourier transform in
Eq.~(\ref{General}) we find the stationary probability distribution
(SPD) of the particle coordinate

\begin{widetext}
\vspace{0.2cm}
\begin{equation}
P_{st}\left( x\right) = \frac{\beta }{\pi }\sum_{l=0}^{\left[
\left( m-1\right) /2\right] }A_{l}\frac{x^{2}\cos \left[ \frac{\pi
\left( m-2l-1\right) }{2m-1}+\varphi _{l}\right] +\beta ^{2}\cos
\left[ \frac{\pi \left( m-2l-1\right) }{2m-1}-\varphi _{l}\right]
}{x^{4}-2x^{2}\beta ^{2}\cos \frac {\pi \left( 4l+1\right) }{2m-1}
+\beta ^{4}}\,.
\label{SPD}
\end{equation}
\vspace{0.2cm}
\end{widetext}

\noindent The parabolic potential profile $U\left( x\right) = \gamma
x^{2}/2$ corresponds to a linear system (\ref{Lang}). In this
situation, from Eqs.~(\ref{Constants}) and (\ref{SPD}) we easily
obtain the following obvious result

\begin{equation}
P_{st}\left( x\right) =\frac{\beta }{\pi \left( x^{2}+\beta
^{2}\right) }\,, \label{Linear}
\end{equation}
\vspace{0.1cm}

\noindent that is due to the stability of the Cauchy distribution
(\ref{Linear}), the probabilistic characteristics of driving noise
increments (see Eq.~(\ref{Cauchy})) and Markovian process
$X\left(t\right)$ are similar (see also Ref.~[West \& Seshadri, 1982]).\\
\indent For quartic potential $\left( m=2\right)$, from the set of
equations (\ref{Constants}), we find $A_{0}=2/\sqrt{3}$ and $\varphi
_{0}=\pi /6$. Substituting these parameters in Eq.~(\ref{SPD}) we
obtain

\begin{equation}
P_{st}\left( x\right) =\frac{\beta ^{3}}{\pi \left(
x^{4}-x^{2}\beta ^{2}+\beta ^{4}\right) }\,, \label{Quartic}
\end{equation}

\noindent which coincides, for $\beta = 1$, with the result obtained
in Refs.~[Chechkin \emph{et al.}, 2002a, 2006]. The plots of
stationary probability distributions (\ref{Quartic}) for L\'{e}vy
flights in symmetric quartic potential, for different values of the
parameter $ \beta $, are shown in Fig.~1.

\begin{figure}[tbph]
\begin{center}
\includegraphics[height=4cm,width=5cm]{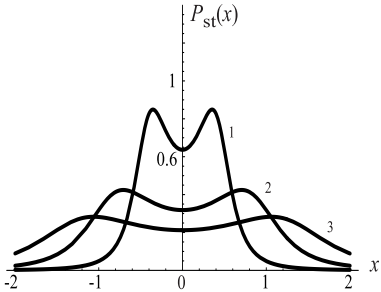}
\end{center}
\caption{{\protect\small \emph{Stationary probability distributions
for L\'{e}vy flights in symmetric quartic potential
$U(x)=\protect\gamma x^{4}/4$ for different values of dimensionless
parameter $\protect\beta$: $(1)~\beta =0.5$, $(2)~\beta =1$,
$(3)~\beta =1.5$.}\protect\bigskip }} \label{beta_1}
\end{figure}

The superdiffusion in the form of L\'{e}vy flight gives rise to a
bimodal stationary probability distribution, when the particle moves
in a monostable potential. This bimodal distribution is a
peculiarity of L\'{e}vy flights. In fact the ordinary diffusion of
the Brownian motion is characterized by unimodal SPD. The SPD of
superdiffusion has two maxima at the points $x=\pm \beta /\sqrt{2}$,
with the value $\left( P_{st}\right) _{max} =4/\left( 3\pi \beta
\right) $. Since the value of the minimum is $P_{st}\left( 0\right)
=1/\left( \pi \beta \right) $, the ratio between maximum and minimum
value is constant and equal to $4/3$. The variance of the particle
coordinate, obtained from Eq.~(\ref{beta_1}) is finite: $\langle
X^2\rangle _{st}=\beta^2$. As a result, the probability distribution
becomes more wide with increasing parameter $\beta
=\sqrt[3]{D/\gamma }$, that is with decreasing the steepness
$\gamma$ of the quartic potential profile, or with increasing the
noise intensity $D$.
\begin{figure}[tbph]
\begin{center}
\includegraphics[height=7cm,width=4cm]{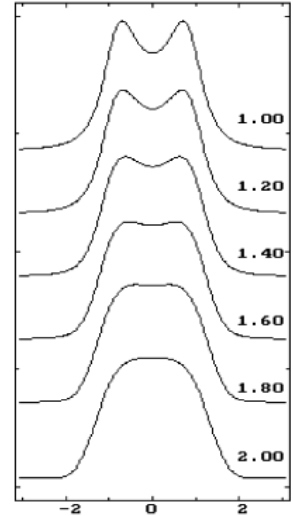}
\end{center}
\caption{{\protect\small \emph{Forms of stationary probability
distribution in the symmetric quartic potential for different
L\'{e}vy indices, from $\alpha =1$ till $\alpha =2$. From
Ref.~[Chechkin et al., 2002a].}}} \label{beta_2}
\end{figure}

\indent A detailed analysis of the solution of the differential
equation~(\ref{Mon-Pot}), for arbitrary L\'{e}vy index $\alpha $ and
quartic potential $(m=2)$ was performed in Refs.~[Chechkin \emph{et
al.}, 2002a, 2004]. In Fig.~2 the profiles of SPD (obtained by an
inverse Fourier transformation) in symmetric quartic potential are
shown for the different L\'{e}vy indices from $\alpha =1$, at the
top of the figure, up to $\alpha =2$ at the bottom~[Chechkin
\emph{et al.}, 2002a]. It is seen that the bimodality is most
strongly expressed for $\alpha =1$ (Cauchy stable noise source). By
increasing the L\'{e}vy index, the bimodal profile smoothes out,
and, finally, it turns into a
unimodal one at $\alpha =2$, recovering the Boltzmann distribution.\\
\indent Carrying out analogous procedure we obtain the stationary
probability distributions for the cases $m=3,4,5$~[Dubkov \&
Spagnolo, 2007]

\begin{widetext}
\begin{eqnarray}
P_{st}\left( x\right) &=& \frac{\beta ^{5}}{\pi \left( x^{2}+\beta
^{2}\right) \left( x^{4}-2\beta ^{2}x^{2}\cos \pi /5+\beta
^{4}\right)}\,, \nonumber\\
P_{st}\left( x\right) &=& \frac{\beta ^{7}}{\pi \left(
x^{4}-2\beta ^{2}x^{2}\cos \pi /7+\beta ^{4}\right) \left(
x^{4}+2\beta ^{2}x^{2}\cos 2\pi /7+\beta ^{4}\right) }\,, \label{Distr}\\
P_{st}\left( x\right) &=& \frac{\beta ^{9}}{\pi \left( x^{2}+\beta
^{2}\right) \left( x^{4}-2\beta ^{2}x^{2}\cos \pi /9+\beta
^{4}\right) \left( x^{4}+2\beta ^{2}x^{2}\cos 4\pi /9+\beta
^{4}\right) }\,. \nonumber
\end{eqnarray}
\vspace{0.03cm}
\end{widetext}

The plots of distributions (\ref{Distr}), for different values of
parameter $\beta $, are respectively shown in Figs.~3--5.\\
\indent It must be emphasized that according to Figs.~3--5, these
distributions remain bimodal and have the same tendency with
increasing $\beta $, but the ratio between maximum and minimum
increases with increasing $m$. From Eqs.~(\ref{Quartic}) and
(\ref{Distr}) we see that the second moment of the particle
coordinate is finite for $m \geq 2$, which confirms the inequality
(\ref{conf}). This means that there is a confinement of the particle
motion due to the steep potential profile, even if the particle
moves according to a superdiffusion in the form of L\'{e}vy flights.
The presence of two maxima is a peculiarity of the superdiffusion
motion. Because of the fast diffusion due to L\'{e}vy flights, the
particle reaches very quickly regions near the potential walls on
the left or on the right with respect to the origin $x = 0$. Then
the particle diffuses around this position, until a
\begin{figure}[tbph]
\begin{center}
\includegraphics[height=5cm,width=6cm]{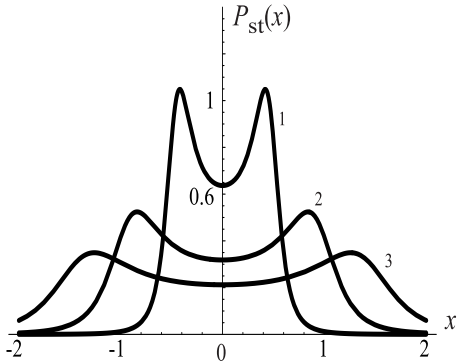}
\end{center}
\caption{{\protect\small \emph{Stationary probability distributions
for L\'{e}vy flights in symmetric potential $U(x)=\protect\gamma
x^{6}/6$ for different values of dimensionless parameter
$\protect\beta $: $(1)~\beta =0.5$, $(2)~\beta =1$, $(3)~\beta
=1.5$.}}}
\label{beta_3}
\end{figure}
\begin{figure}[tbph]
\begin{center}
\includegraphics[height=5cm,width=6cm]{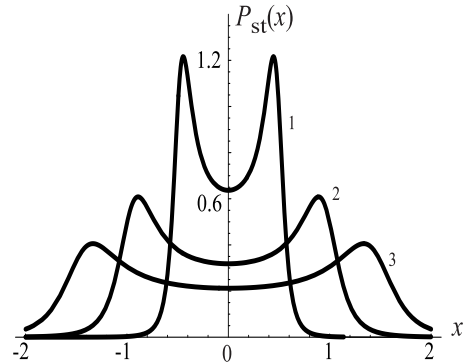}
\end{center}
\caption{{\protect\small \emph{Stationary probability distributions
for L\'{e}vy flights in symmetric potential $U(x)=\protect\gamma
x^{8}/8$ for different values of dimensionless parameter
$\protect\beta $: $(1)~\beta =0.5$, $(2)~\beta =1$, $(3)~\beta
=1.5$.}}}
\label{beta_4}
\end{figure}
new flight moves it in the opposite direction to reach the other
potential wall. As a result, the particle spends a large time in
some
\begin{figure}[tbph]
\begin{center}
\includegraphics[height=5cm,width=6cm]{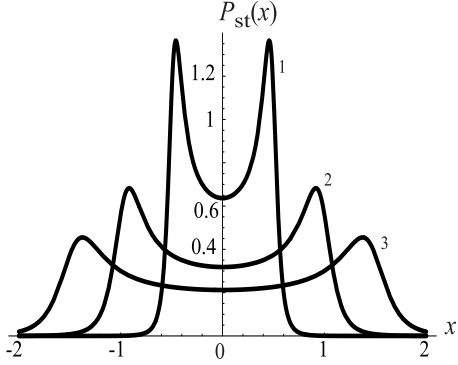}
\end{center}
\caption{{\protect\small \emph{Stationary probability distributions
for L\'{e}vy flights in symmetric potential $U(x)=\protect\gamma
x^{10}/10$ for different values of dimensionless parameter
$\protect\beta $: $(1)~\beta =0.5$, $(2)~\beta =1$, $(3)~\beta
=1.5$.} \protect\bigskip }}
\label{beta_5}
\end{figure}
symmetric areas with respect to the point $x=0$, differently from
the Brownian diffusion in monostable potential profiles. These
symmetric areas lie near the maxima of the bimodal SPD. For fixed
$D$ and $m$, these maxima are closer or far away the point $x = 0$
depending on the greater or smaller steepness $\gamma$ of the
potential profile. This corresponds to a greater or smaller
confinement of the particle motion. Of course, such confinement is
more pronounced for greater $m$, that is for steeper potential
profiles.\\
\indent On the basis of Eqs.~(\ref{Linear})--(\ref{Distr}) and the
known behavior of density tails (\ref{Tail}), we can write the
general expressions for stationary probability distribution in the
case of potential $U\left( x\right) =\gamma x^{2m}/\left( 2m\right)
$ with odd $m=2n+1$~[Dubkov \& Spagnolo, 2007]
\begin{widetext}
\vspace{0.1cm}
\begin{equation}
P_{st}\left( x\right) =\frac{\beta ^{4n+1}}{\pi \left( x^{2}+\beta
^{2}\right) }\prod\limits_{l=0}^{n-1}\frac{1}{x^{4}-2\beta
^{2}x^{2}\cos \left[ \pi \left( 4l+1\right) /\left( 4n+1\right)
\right] +\beta ^{4}}\,,\label{Final-1}
\end{equation}
\vspace{0.1cm}
\end{widetext}

\noindent and even $m=2n$

\begin{widetext}
\vspace{0.1cm}
\begin{equation}
P_{st}\left( x\right) =\frac{\beta ^{4n-1}}{\pi
}\prod\limits_{l=0}^{n-1} \frac{1}{x^{4}-2\beta ^{2}x^{2}\cos \left[
\pi \left( 4l+1\right) /\left( 4n-1\right) \right] +\beta ^{4}}\, .
\label{Final-2}
\end{equation}
\vspace{0.1cm}
\end{widetext}

The strong proof of non-unimodality of the SPD for symmetric
monostable potential $U\left( x\right) =\left\vert x\right\vert^c/c$
in the case $c>2$ was given in Ref.~[Chechkin \emph{et al.}, 2004].
Indeed, from Eq.~(\ref{FFPE}) we have for SPD

\begin{equation}
\frac{d}{dx}\left[ \left\vert x\right\vert^{c-1}\mathrm{sgn}~x\cdot
P_{st}\right] +D\frac{d^{\alpha}P_{st}}{d\left\vert x\right\vert
^{\alpha}}=0.\label{FFPE-st}
\end{equation}

\noindent As a result, from Eqs.~(\ref{FFPE-st}) and (60) at the
point $x=0$ we obtain

\begin{equation}
\left. \frac{d^{\alpha}P_{st}}{d\left\vert x\right\vert
^{\alpha}}\right| _{x=0}=\int_{-\infty}^{+\infty}\frac{P_{st}\left(
-z\right) -P_{st}\left( 0\right) }{\left\vert z\right\vert
^{\alpha+1}}=0. \label{zero}
\end{equation}

\noindent Because of the symmetry of the SPD $P_{st}\left(
x\right)$, Eq.~(\ref{zero}) gives

\begin{equation}
\int_{0}^{\infty}\frac{P_{st}\left( z\right) -P_{st}\left(
0\right) }{z^{\alpha+1}}=0. \label{proof}
\end{equation}

\noindent For unimodal probability distribution with the maximum at
the origin, the integral in the left side of Eq.~(\ref{proof})
should
be negative, which contradicts Eq.~(\ref{proof}).\\
\indent The estimation of bifurcation time for transition from
unimodal initial distribution to bimodal stationary one for the
quartic potential $(c=4)$ was done in Refs.~[Chechkin \emph{et al.},
2004, 2006]. The dependence of this bufurcation time $t_{12}$ from
L\'{e}vy index $\alpha $ is plotted in Fig.~6.
\begin{figure}[tbph]
\begin{center}
\includegraphics[height=5cm,width=6cm]{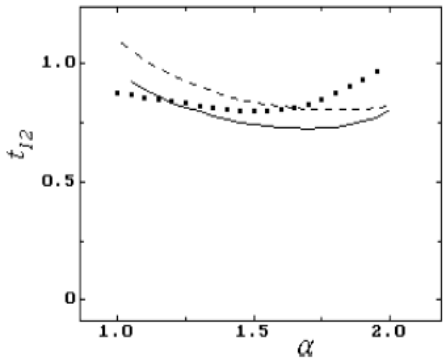}
\end{center}
\caption{{\protect\small \emph{Bifurcation time $t_{12}$ versus
L\'{e}vy index $\alpha $ for quartic potential. Black dots:
bifurcation time deduced from the numerical solution of the
fractional Fokker–-Planck equation. Dashed and solid lines: two
subsequent approximations. From Refs.~[Chechkin et al., 2004,
2006].}}} \label{beta_6}
\end{figure}
Also authors proved an existence of a transient trimodal state
between initial unimodal and final bimodal ones. This evolution,
shown in Fig.~7, can be only observed for monostable potential with
$c>4$ and for fixed values of the L\'{e}vy index $\alpha$.
\begin{figure}[tbph]
\begin{center}
\includegraphics[height=7cm,width=4cm]{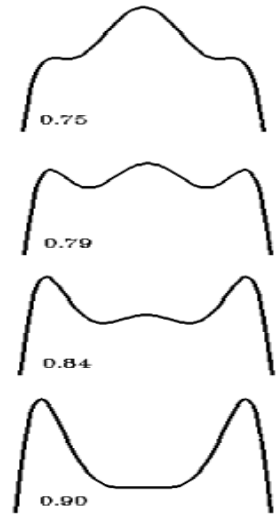}
\end{center}
\caption{{\protect\small \emph{The evolution of the probability
distribution for $\alpha = 1.2$ and $c=5.5$ from unimodal through
trimodal to bimodal. From Refs.~[Chechkin et al., 2004, 2006].}}}
\label{beta_7}
\end{figure}
The corresponding bifurcation times of transitions $t_{13}$
(unimodal $\rightarrow$ trimodal) and $t_{32}$ (trimodal
$\rightarrow$ bimodal) versus L\'{e}vy index $\alpha $, with
potential exponent $c=5.5$, are plotted in Fig.~8.
\begin{figure}[tbph]
\begin{center}
\includegraphics[height=6cm,width=7cm]{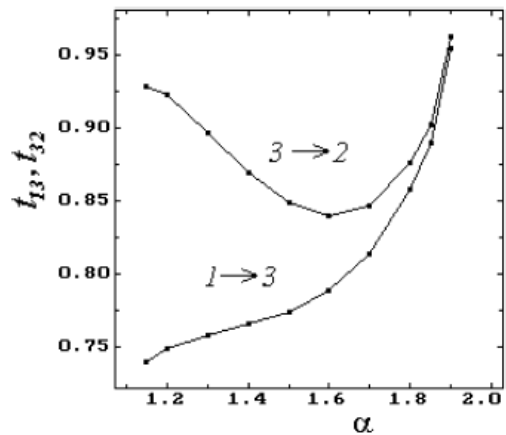}
\end{center}
\caption{{\protect\small \emph{Bifurcation times $t_{13}$ and
$t_{32}$ versus L\'{e}vy index $\alpha $ for the power potential
with the exponent $c=5.5$. From Ref.~[Chechkin \emph{et al.},
2004].}}}
\label{beta_8}
\end{figure}

\section{Barrier crossing}

The problem of escape from metastable states investigated by Kramers
[Kramers, 1940] is ubiquitous in almost all scientific
areas~[H\"{a}nggi \emph{et al.}, 1990; Spagnolo \emph{et al.},
2007]. Since many stochastic processes do not obey the Central Limit
Theorem, the corresponding Kramers escape behavior will differ. An
interesting example is given by the $\alpha$-stable noise-induced
barrier crossing in long paleoclimatic time series~[Ditlevsen,
1999a]. Another new application is the escape from traps in
optical or plasma systems~[Fajans \& Schmidt, 2004].\\
\indent The main tools to investigate the barrier crossing problem
for L\'{e}vy flights are the first passage times, crossing times,
arrival time and residence times. We should emphasize that the
problem of mean first passage time (MFPT) meets with some
difficulties because free L\'{e}vy flights represent a special class
of discontinuous Markovian processes with infinite mean squared
displacement. First of all, the fractional Fokker-Planck
equation~(\ref{FFPE}) is integro-differential, and the conditions at
absorbing and reflecting boundaries differ from the usual conditions
for ordinary diffusion. Superdiffusion motion is characterized by
the presence of jumps, and, as a result, a particle can reach
instantaneously the boundary from arbitrary position. One can
mention some erroneous results for L\'{e}vy flights obtained in
Ref.~[Gitterman, 2000], because author used the traditional
conditions at two absorbing boundaries (see the related
correspondence~[Yuste \& Lindenberg, 2004; Gitterman, 2004]). The
numerical results for the first passage time of free L\'{e}vy
flights confined in a finite interval were presented in Ref.~[Dybiec
\emph{et al.}, 2006]. The complexity of the first passage time
statistics (mean first passage time, cumulative first passage time
distribution) was elucidated together with a discussion of the
proper setup of corresponding boundary conditions, that correctly
yield the statistics of first passages for these non-Gaussian
noises. In particular, it has been demonstrated by numerical studies
that the use of the local boundary condition of vanishing
probability flux in the case of reflection, and vanishing
probability in the case of absorbtion, valid for normal Brownian
motion, no longer apply for L\'{e}vy flights. This in turn requires
the use of nonlocal boundary conditions. Dybiec with co-authors
found a nonmonotonic behavior of the MFPT for two absorbing
boundaries, with the maximum being assumed for $\alpha=1$ (see
Fig.~9), in contrast with a monotonic increase for reflecting and
absorbing boundaries.
\begin{figure}[tbph]
\begin{center}
\includegraphics[height=7cm,width=5cm]{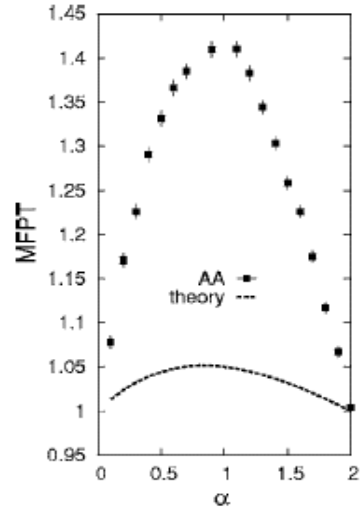}
\end{center}
\caption{{\protect\small \emph{Mean first passage time versus
L\'{e}vy index $\alpha $ of confined motion between two absorbing
boundaries driven by stable symmetric L\'{e}vy white noise. From
Ref.~[Dybiec \emph{et al.}, 2006].} \protect\bigskip }}
\label{beta_9}
\end{figure}

\indent According to the Kramers law, the probability distribution
of the escape time from a potential well with the barrier of height
$U_{0}$, has the exponential form

\begin{equation}
p\left( t\right) =\frac{1}{T_{c}}\exp\left\{
-\frac{t}{T_{c}}\right\} \label{ABV-01}
\end{equation}

\noindent with mean crossing time

\begin{equation}
T_{c}=C\exp\left\{ \frac{U_{0}}{D}\right\} ,\label{ABV-02}
\end{equation}

\noindent where $C$ is some positive prefactor and $D$ is the noise
intensity. The problem how the stable nature of L\'{e}vy flight
processes generalizes the barrier crossing behavior of the classical
Kramers problem was investigated, both numerically and analytically,
in Ref.~[Chechkin \emph{et al.}, 2003b, 2005, 2006, 2007]. Authors
considered L\'{e}vy flights in a bistable potential $U\left(
x\right) $ by numerical solution of Eq.~(\ref{LSN}). It was shown
that although the survival probability decays again exponentially as
in Eq.~(\ref{ABV-01}), the mean escape time $T_{c}$ has a power-law
dependence on the noise intensity $D$

\begin{equation}
T_{c}\simeq\frac{C(\alpha)}{D^{\mu(\alpha)}}\,,\label{ABV-03}
\end{equation}

\noindent where the prefactor $C$ and the exponent $\mu$ depend on
the L\'{e}vy index $\alpha$. Using the Fourier transform, i.e.
Eq.~(\ref{Char-1}), the mean escape rate was found for large values
of $1/D$ in the case of Cauchy stable noise $\left( \alpha=1\right)
$ in the framework of the constant flux approximation across the
barrier. The probability law and the mean value of escape time from
a potential well for all values of the stability index
$\alpha\in(0,2)$, in the limit of small L\'{e}vy driving noise, were
also determined in the paper~[Imkeller \& Pavlyukevich, 2006] by
purely probabilistic methods. Escape times had the same exponential
distribution~(\ref{ABV-01}), and the mean value depends on the noise
intensity $D$ in accordance with Eq.~(\ref{ABV-03}) with
$\mu(\alpha) = 1$ and pre-factor $C$ depending on $\alpha$ and the
distance between the local extrema
of the potential.\\
\indent The barrier crossing of a particle driven by symmetric
L\'{e}vy noise of index $\alpha$ and intensity $D$ for three
different generic types of potentials was numerically investigated
in Ref.~[Chechkin \emph{et al.}, 2007]. Specifically: (i) a bistable
potential, (ii) a metastable potential, and (iii) a truncated
harmonic potential, were considered. For the low noise intensity
regime, the result of Eq.~(\ref{ABV-03}) was recovered. As it was
shown, the exponent $\mu(\alpha)$ remains approximately constant,
$\mu\approx1$ for $0<\alpha<2$; at $\alpha=2$ the power-law form of
$T_{c}$ changes into the exponential dependence~(\ref{ABV-02}). It
exhibits a divergence-like behavior as $\alpha$ approaches $2$. In
this regime a monotonous increase of the escape time $T_{c}$ with
increasing $\alpha$ (keeping the noise intensity $D$ constant) was
observed. For low noise intensities the escape times correspond to
barrier crossing by multiple L\'{e}vy steps. For high noise
intensities, the escape time curves collapse for all values of
$\alpha$. At intermediate noise intensities, the escape time
exhibits non-monotonic dependence on the index $\alpha$ as in
Fig.~9, while still retains the exponential form of the escape time density.\\
\indent The first arrival time is an appropriate parameter to
analyze the barrier crossing problem for L\'{e}vy flights. If we
insert in fractional Fokker-Planck equation (\ref{FFPE}) a
$\delta$-sink of strength $q\left( t\right) $ in the origin, we
obtain the following equation for the non-normalized probability
density function $P\left( x,t\right) $

\begin{equation}
\frac{\partial P}{\partial t}=\frac{\partial}{\partial x}\left[
U^{\prime }\left( x\right) P\right]
+D\frac{\partial^{\alpha}P}{\partial\left\vert x\right\vert
^{\alpha}}-q\left( t\right) \delta\left( x\right) ,\label{ABV-04}
\end{equation}
\vspace{0.03cm}

\noindent from which by integration over all space we may define the
quantity

\begin{equation}
q\left( t\right) =-\frac{d}{dt}\int_{-\infty}^{+\infty}P\left(
x,t\right) dx,\label{ABV-05}
\end{equation}

\noindent which is the negative time derivative of the survival
probability. According to definition (\ref{ABV-05}), $q\left(
t\right) $ represents the probability density function of \emph{the
first arrival time}: once a random walker arrives at the sink it is
annihilated. As it was shown in the paper [Chechkin \emph{et al.},
2003b] for free L\'{e}vy flights $\left( U\left( x\right)
=0\right)$, the first arrival time distribution has a heavy tail

\begin{equation}
q\left( t\right) \sim t^{1/\alpha-2}\label{ABV-06}
\end{equation}

\noindent with exponent depending on L\'{e}vy index $\alpha$ $\left(
1<\alpha<2\right) $ and differing from universal Sparre Andersen
result [Sparre Andersen, 1953, 1954] for the probability density
function of first passage time for arbitrary Markovian process

\begin{equation}
p\left( t\right) \sim t^{-3/2}.\label{ABV-07}
\end{equation}

In the Gaussian case $\left( \alpha = 2\right)$, the quantity
(\ref{ABV-06}) is equivalent to the first passage time probability
density (\ref{ABV-07}). From a random walk perspective, this is due
to the fact that individual steps are of the same increment, and the
jump length statistics therefore ensures that the walker cannot hop
across the sink in a long jump without actually hitting the sink and
being absorbed. This behavior becomes drastically different for
L\'{e}vy jump length statistics: there, the particle can easily
cross the sink in a long jump. Thus, before eventually being
absorbed, it can pass by the sink location numerous times, and
therefore the statistics of the first arrival will be different from
that of the first passage. The result (\ref{ABV-07}) for L\'{e}vy
flights was also confirmed numerically
in the paper [Koren \emph{et al.}, 2007].\\
\indent At last, the nonlinear relaxation time technique is also
suitable for investigations of L\'{e}vy flights temporal
characteristics. According to definition, the mean residence time
in the interval $\left( L_{1},L_{2}\right) $ reads
\begin{equation}
T\left( x_{0}\right) =\int_{0}^{\infty}dt
\int_{L_{1}}^{L_{2}}P\left(\left. x,t\right\vert x_{0},0\right)
dx,\label{ABV-08}
\end{equation}
\vspace{0.01cm}

\noindent where $x_{0}$ is the initial position of all particles
$\left( x_{0}\in\left( L_{1},L_{2}\right) \right) $ and $P\left(
\left. x,t\right\vert x_{0},0\right)$ is the probability density of
transitions. We do not need to think about the boundary conditions
in this case because we are concerned with the overall time spent by
the particle in the fixed interval. Changing the order of
integration in Eq.~(\ref{ABV-08}) we obtain
\begin{equation}
T\left( x_{0}\right) =\int_{L_{1}}^{L_{2}}Y\left( x,x_{0}
,0\right) dx,\label{ABV-09}
\end{equation}
where $Y\left( x,x_{0},s\right) $ is the Laplace transform of the
transient probability density $P\left( \left. x,t\right\vert
x_{0},0\right)$
\begin{equation}
Y\left( x,x_{0},s\right) =\int_{0}^{\infty}P\left( \left.
x,t\right\vert x_{0},0\right) e^{-st}dt.
\end{equation}
\vspace{0.001cm}

\noindent Making the Laplace transform in the fractional
Fokker-Planck equation (\ref{FFPE}) and taking into account the
initial condition $P\left( \left. x,0\right\vert x_{0},0\right)
=\delta\left( x-x_{0}\right)$, we get
\begin{equation}
\frac{d}{dx}\left[ U^{\prime}\left( x\right)Y\right]
+D\frac{d^{\alpha} Y}{d\left\vert x\right\vert
^{\alpha}}-sY=-\delta\left( x-x_{0}\right) .\label{ABV-10}
\end{equation}
\vspace{0.01cm}

\noindent If we put $s=0$ in Eq.~(\ref{ABV-10}) and make the Fourier
transform we obtain
\begin{equation}
U^{\prime}\left( -i\frac{d}{dk}\right) \widetilde{Y}-iD\left\vert
k\right\vert ^{\alpha-1}\mathrm{sgn}\left( k\right)
\widetilde{Y}=\frac{e^{ikx_{0}}}{ik},\label{ABV-11}
\end{equation}

\noindent where
\begin{equation}
\widetilde{Y}\left( k,x_{0}\right) = \int_{-\infty}^{+\infty}
Y\left( x,x_{0},0\right) e^{+ikx}dx.
\label{ABV-12}
\end{equation}
\vspace{0.01cm}

\noindent After solving Eq.~(\ref{ABV-11}) we can calculate the mean
residence time as (see Eqs.~(\ref{ABV-09}) and (\ref{ABV-12}))
\begin{equation}
T\left( x_{0}\right) =\frac{1}{2\pi i}\int_{-\infty}^{+\infty}
\frac{e^{-ikL_{1}}-e^{-ikL_{2}}}{k}\widetilde{Y}\left(k,x_{0}\right)
dk.
\label{ABV-13}
\end{equation}
\vspace{0.01cm}

\noindent Equations (\ref{ABV-11}) and (\ref{ABV-13}) are useful
tools to analyze the temporal characteristics of L\'{e}vy flights in
different potential profiles $U\left( x\right)$.

\section{Conclusions}
\vspace{-0.009cm}

 In this tutorial paper, after some short historical notes on
normal diffusion and superdiffusion, we introduce the L\'{e}vy
flights as self-similar L\'{e}vy processes. After the definition of
the strictly stable random variables, the subfamily of the L\'{e}vy
motion is introduced with the fractional differential equation for
L\'{e}vy flight superdiffusion. We used then functional analysis
approach to derive the fractional Fokker-Planck equation directly
from Langevin equation with symmetric $\alpha$-stable L\'{e}vy
noise. This approach allows to describe anomalous diffusion in the
form of L\'{e}vy flights. We obtained the general formula for
stationary probability distribution of superdiffusion in symmetric
smooth monostable potential for Cauchy driving noise. All
distributions have bimodal shape and become more narrow with
increasing steepness of the potential or with decreasing noise
intensity. We found that the variance of the particle coordinate is
finite for quartic potential profile and for steeper potential
profiles, that is a confinement of the particle in a superdiffusion
motion in the form of L\'{e}vy flights. As a result, we can evaluate
the power spectral density of a stationary motion. We have also
discussed recently obtained analytical and numerical results for
time characteristics of L\'{e}vy flights. Special attention was
given for some difficulties with formulation of the correct boundary
conditions for mean first passage time problem. As it was shown, the
arrival and residence times are more appropriate characteristics for
investigations of L\'{e}vy flights in different potential profiles.

\vspace{0.001cm} \noindent \textbf{Acknowledgments}

\noindent This work has been supported by MIUR, CNISM, and by
Russian Foundation for Basic Research (projects 07-01-00517 and
08-02-01259).

\end{document}